\documentclass[pdflatex,sn-mathphys-num]{sn-jnl}

\usepackage{graphicx}%
\usepackage{multirow}%
\usepackage{amsmath,amssymb,amsfonts}%
\usepackage{amsthm}%
\usepackage{mathrsfs}%
\usepackage[title]{appendix}%
\usepackage{xcolor}%
\usepackage{textcomp}%
\usepackage{manyfoot}%
\usepackage{booktabs}%
\usepackage{algorithm}%
\usepackage{algorithmicx}%
\usepackage{algpseudocode}%
\usepackage{listings}%

\usepackage{tikz}%
\usetikzlibrary{quantikz2}%
\usepackage{subfig}%

\usepackage{comment}

\theoremstyle{thmstyleone}%
\theoremstyle{thmstyletwo}%

\theoremstyle{thmstylethree}%

\begin{document}

\title[Article Title]{On the Convergence of Markov Chain Distribution within Quantum Walk Circuit Subspace}

\author*[1]{\fnm{Aingeru} \sur{Ramos}}\email{aingeru.ramos@bcmaterials.net}

\author[2]{\fnm{Jose A.} \sur{Pascual}}\email{joseantonio.pascual@ehu.eus}
\equalcont{These authors contributed equally to this work.}

\author[2]{\fnm{Javier} \sur{Navaridas}}\email{javier.navaridas@ehu.eus}
\equalcont{These authors contributed equally to this work.}

\author[3]{\fnm{Ivan} \sur{Coluzza}}\email{ic33@rice.edu}
\equalcont{These authors contributed equally to this work.}

\affil*[1]{\orgname{Basque Center for Materials, Applications
and Nanostructures (BCMaterials)}, \orgaddress{\street{Martina Casiano building. EHU Technology Park.}, \city{Leioa}, \postcode{48940}, \country{Spain}}}

\affil[2]{\orgdiv{Dept. of Computer Architecture and Technology}, \orgname{University of the Basque Country (UPV/EHU)}, \orgaddress{\street{Manuel Lardizabal Pasalekua},\\ \city{San Sebastián}, \postcode{20018}, \country{Spain}}}

\affil[3]{\orgdiv{Department of Chemistry, Ken Kennedy Institute, Center for Theoretical Biological Physics and Smalley-Curl Institute}, \orgname{Rice University}, \orgaddress{\street{6100 Main St}, \city{Houston}, \postcode{TX 77005}, \country{USA}}}


\abstract{Markov Chain Monte Carlo (MCMC) methods are algorithms for sampling probability distributions, commonly applied to the Boltzmann distribution in physical and chemical models such as protein folding and the Ising model. These methods enable exploration of such systems by sampling their most probable states. However, sampling multidimensional and multimodal distributions with MCMC requires substantial computational resources, leading to the development of techniques aimed at improving sampling efficiency. In this context, quantum computing, with its potential to accelerate classical methods, emerges as a promising solution to the sampling problem. In this work, we present the design of a new circuit based on the Discrete Quantum Walk (DQW) algorithm to perform MCMC sampling over a desired distributions. Simulation results show convergence behavior in the superposition of the quantum register that encodes the target distribution. This design is further refined to increase convergence speed and, consequently, the scalability of the algorithm.}



\maketitle

\section{Introduction}
\label{sec:intro}

The study of physical systems requires solving equations derived from the natural laws that govern them, often an intractable task due to (i) the lack of a closed-form solution, or (ii) the complexity of the equations, which prevents analytical resolution~\cite{frenkel}. For this reason, the use of computational methods has become fundamental in physics, and in general, across all scientific disciplines. The use of such methods dates back to the 20th century and marks the birth of computational physics and its most fundamental tool: the simulation of physical models.

These simulations can be broadly divided into two categories. The first one encompasses physical simulations, which create a virtual version of the system under study and emulate the physical laws governing its evolution. Due to the complexity of these systems, scientists tend to rely on simplified models to save computational resources and obtain results more rapidly, often at the expense of accuracy. On the other hand, numerical simulations do not emulate the system under study, instead, provide approximate solutions using numerical and statistical methods. This enables the extraction of macroscopic characteristics such as pressure, temperature, or density, parameters relevant for analyzing and understanding the system under study. In this context, Markov Chain Monte Carlo (MCMC) methods have become widely used. The first algorithm developed in this family of methods was the Metropolis-Hastings (MH) algorithm, based on the works of Metropolis~\cite{metropolis} and Hastings~\cite{hastings}, which defines the general scheme of all MCMC processes. These algorithms, capable of sampling any given probability distribution, are commonly used in physical-chemical simulations together with the Boltzmann distribution which describes the probability of a system being in a particular state. As a result, sampling the Boltzmann distribution through MCMC methods enables exploration of the system’s state space and, yields samples of its most probable states. However, in real systems, with immense and often multidimensional state spaces, the effectiveness of these methods decreases, often resulting in non-representative samples of the target distribution. Various techniques have been developed to accelerate convergence, but all of them, in one way or another, demand a substantial increase in computational resources.

Quantum computing is an emerging field that explores the use of quantum-mechanical principles to perform computation~\cite{nielsen}. Unlike classical computers, quantum computers encode information in qubits, which can represent multiple states simultaneously, enabling calculations across multiple states at once. This property, under specific conditions, can produce a quadratic acceleration in the performance of algorithms, offering great potential in applications that require huge computational resources, such as molecular simulation, artificial intelligence, and optimization. However, quantum computing still faces significant technical challenges, such as quantum error correction and the stability of qubits in superposition~\cite{preskill}. Despite these obstacles, the immense potential of the technology continues to drive advancements in both algorithm design and quantum hardware development. 

In this paper, we develop a quantum circuit based on the Discrete Quantum Walk (DQW) scheme, aiming for a quantum implementation of the MCMC algorithm for distribution sampling. We refer to this circuit as Quantum MCMC, or simply QMCMC. Our results indicate that the proposed circuit is capable of capturing the target distribution. Then, the design is expanded to boost initial convergence speed by increasing the number of movement evaluations per iteration.

\subsection{Related Work}
\label{sec:related}

The concept of Quantum Walk (QW) was introduced by Aharanov and his collaborators in 1993~\cite{aharonov}, proposing an iterative circuit inspired by the behavior of classical Random Walks. Unlike its classical counterpart, where each iteration only moves in a randomly chosen direction, QWs using superposition can reflect movement in multiple directions in a single iteration. This property allows for faster exploration of the state space, which can be an advantage over classical methods. One of the first results demonstrating an algorithmic advantage of QWs over classical methods was the work of Ambainis, later published in~\cite{ambainis_qw1d}. There, Ambainis analyzed the behavior of QWs in search tasks on graphs of different dimensionalities and demonstrated that while the advantage offered by the circuit is considerably reduced at lower dimensionalities, QWs significantly outperform classical search methods in certain high-dimensional structures. This result marked a turning point in QW research, consolidating them as a promising tool for the design of quantum circuits.

Another important work in the field of QWs is Szegedy’s 2004 contribution~\cite{szegedy}, where he generalized QWs by providing a framework that allows the quantization of reversible Markov chains. His construction is based on a unitary operator formed from reflections, which differs from the original coin-position quantum walk models. This approach not only provides a powerful method to achieve quadratic speedups in key parameters such as mixing and hitting times but also has had significant impact on the development of quantum algorithms based on quantum walks.

In 2009, Childs~\cite{child_universal} showed that QWs are not only useful for accelerating search algorithms, but also have the potential to be adapted for universal quantum computation. This discovery significantly expanded their scope, extending their applications beyond search to the simulation of complex quantum systems and the design of more general-purpose quantum algorithms. Since the pioneering work of Aharonov, Ambainis, and Szegedy, research on QWs has evolved in multiple directions. Their demonstrated capabilities in accelerating search algorithms and simulating quantum dynamics have paved the way for applications in quantum optimization, physical and chemical simulations, and the development of universal quantum algorithms~\cite{zahringer,schreiber,peruzzo}.

Regarding the creation of quantum sampling methods, the community has explored various approaches. Among the most relevant is the Quantum Approximate Optimization Algorithm (QAOA), which, although originally conceived as a method for solving combinatorial problems, generates a probabilistic output distribution that can be used as a sampling method~\cite{qaoa_betterconvergence,qaoa_customoperator,qaoa_thermaldistro}. Likewise, the use of the Quantum Annealing (QA) computing scheme has been explored. Although QA is a different paradigm from circuit-based quantum computing, it has proven particularly useful in sampling problems due to the adiabatic evolution that characterizes these systems, which allows relevant probability distributions to be approximated~\cite{qa_dwave,qa_mcmc}. Finally, hybrid methods have also been explored that combine the use of quantum methods and classical processing, “quantizing” those algorithm processes that can be most difficult for classical computers~\cite{hybrid_mcmc,hybrid_moremcmc}.

However, in Quantum Walks research has mainly focused on developing circuits for graph search and studying how they propagate through state space. Although the Szegedy circuit reproduces the dynamics of classical Markov chains, the need to prepare an initial state strongly correlated with the target distribution impedes its use as a sampling method such as a Markov Chain Monte Carlo (MCMC) algorithm. Subsequent work on the Szegedy circuit has focused on the implementation and optimization of the operators defined by Szegedy for execution on real hardware~\cite{tucci}.

In this work, we return to Aharonov's and Ambainis's Quantum Walk scheme. Starting from there, we add new operators and qregisters that allow the circuit to follow the algorithmic steps of the Metropolis-Hastings algorithm and force the distribution of positions to fit the desired shape, described by the function $ f $ that we want to sample.

\section{Theoretical Framework}
\label{sec:theo_frame}

\subsection{Markov Chain Monte Carlo}
\label{subsec:mcmc}

Let $ X $ be a random variable that follows a probability distribution $ f(x) $, and consider a random sample $ \{X_0, X_1, X_2, ..., X_n\} $ of size $ n+1 $, where the frequency of each value of X is proportional to the probability given by $ f $. While some distributions have well-known direct sampling methods, these are rare. In most cases, approximation methods are required, such as {\bf Monte Carlo methods}, which provide independent samples from the distribution. In contrast, MCMC methods also sample from any target distribution $f$, but the resulting samples are dependent, forming a chain constructed by the Markov process.

The basic scheme of the Metropolis-Hastings algorithm is as follows. Let $ x $ be a variable initialized with a random value $ x_{init} $. For each of the $ N $ iterations, the algorithm performs three steps: (i) a candidate trial $ x_{trial} $ is generated using a generator function $g$ such that $ x_{trial} = g(x) $, (ii) the acceptance probability for the new state is computed using the acceptance function $A$, as shown in Equation \eqref{eq:acc_f}, and (iii) the current value of $ x $ is either replaced by $ x_{trial} $ according to the previous acceptance probability, or kept and stored in a list. After $ N $ iterations, the list contains a Markov chain that approximates the target distribution $f$. It is noteworthy that if the generator function $ g $ is symmetrical, the expression $ g(x_{\text{trial}} |x) = g(x |x_{\text{trial}}) $ holds; and consequently, the Equation \eqref{eq:acc_f} defining the probability of acceptance is greatly simplified, depending exclusively on the relation in $ f $ between the current state $ x $ and the trial $ x_{\text{trial}} $.

\begin{eqnarray}
\label{eq:acc_f}
\centering
\begin{aligned}
    A(x_{trial}, x) = min\left(1, \frac{f(x_{trial})\cdot g(x | x_{trial})}{f(x)\cdot g(x_{trial} | x)}\right)
\end{aligned}
\end{eqnarray}

\begin{figure}[htbp]
\centering
    \includegraphics[width=0.618\linewidth]{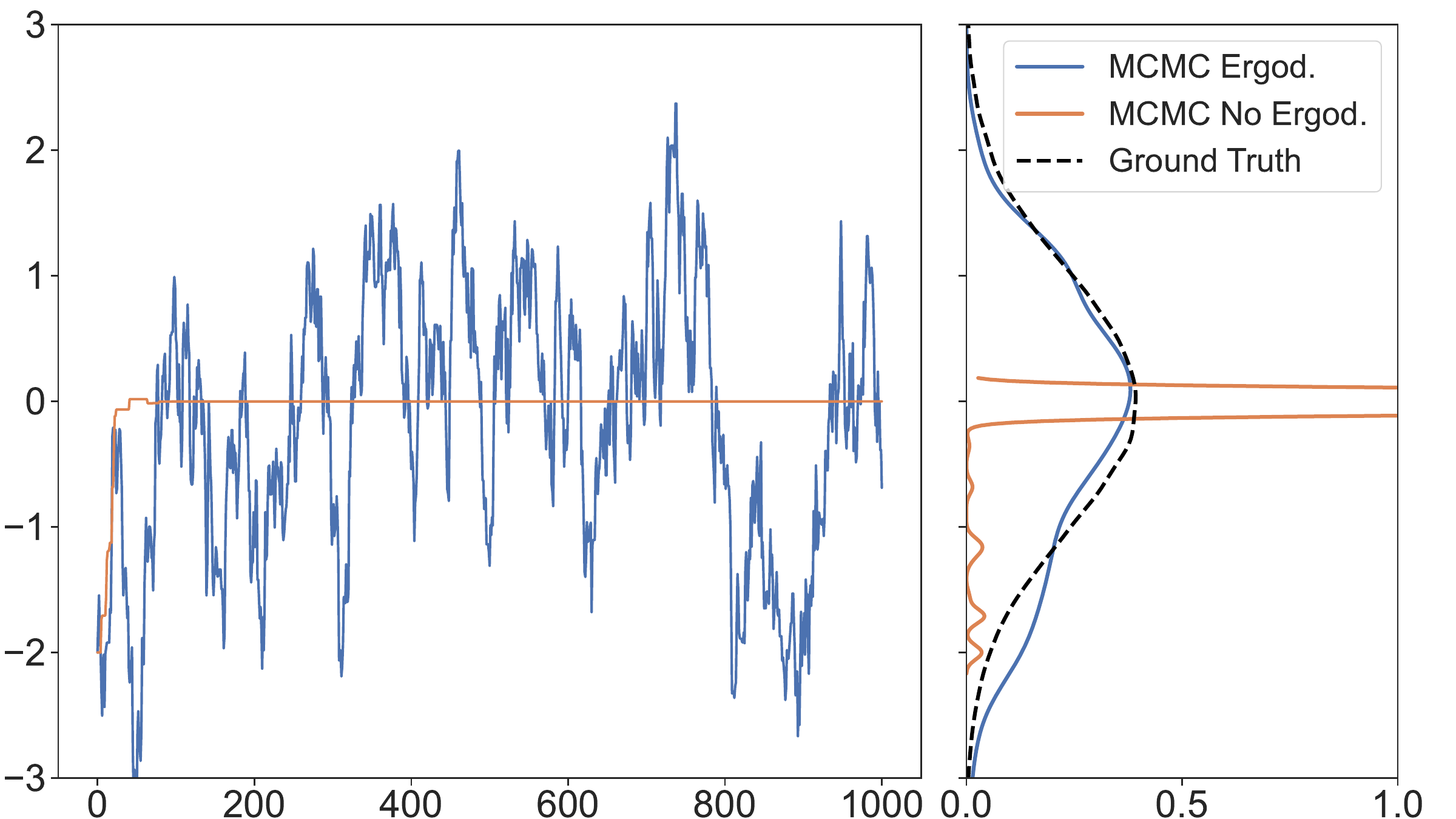}
    \vspace{0.05cm}
    \caption{Left: sample chains generated using different acceptance functions. Right: the resulting distributions for each chain, compared against the true distribution \({\mathcal{N}}(0,1)\). The figure illustrates how ergodicity affects convergence to the target distribution.}

\label{fig:accp_cmp} 
\end{figure}

All algorithms in the MCMC family follow this basic scheme, with slight variations in the proposal or acceptance functions. However, they must preserve ergodicity, a key property that ensures, given enough iterations, the algorithm can reach all possible states of the target distribution $f$. This property depends on the design of the proposal function $g$, the acceptance function $A$, and the distribution $f$ itself. When ergodicity holds, the process is guaranteed to eventually explore the entire support of the distribution as $ N \rightarrow \infty $, producing statistically representative samples (see Figure~\ref{fig:accp_cmp}).

Despite their utility, MCMC algorithms face a significant limitation: slow convergence, particularly when the distribution $f$ is multimodal, high-dimensional, or defined over a large or intractable state space. In such cases, the number of iterations required to obtain representative samples can be prohibitively large, making the algorithm impractical in its basic form. Although many improved variants\cite{gibbs}\cite{pt_theory}\cite{pt_deri_adaptativept} and convergence-acceleration techniques have been proposed, the challenge of efficient and scalable sampling remains an open research problem.

\subsection{Discrete Quantum Walk}
\label{subsec:dqw}

\subsubsection{Quantum States and Gates}
\label{subsubsec:qstates_qgates}

The state of a quantum system is described by a column vector $ \ket{\psi} $ in a Hilbert space $ \mathcal{H} $, which is spanned by a set of orthonormal basis vectors $ \{ \ket{\phi_0}, \ket{\phi_1}, \ldots, \ket{\phi_{n\text{-}1}} \} $, each in the complex vector space $ {\mathbb{C}}^n $. In quantum mechanics, the basis states $ \ket{\phi_i} $ correspond to measurable, distinguishable outcomes, while a general state $ \ket{\psi} $ is a linear combination (superposition) of the basis vectors\footnote{The symbol $ \bra{\psi} $ represents the transposed conjugate of $ \ket{\psi} $ ($ \bra{\psi} = \ket{\psi}^\dagger) $. That is, $ \bra{\psi} $ is the row vector of the same $ \ket{\psi} $ column vector.}. The coefficients $c_i$ represent the probability amplitudes of measuring each $ \ket{\phi_i} $ state, so they must fulfill the normalization condition. These superposition states are a fundamental feature of quantum systems and form the basis for quantum parallelism.

\begin{eqnarray*}
\begin{aligned}
    \ket{\psi} = \sum_{i=0}^{|{\mathcal{H}}|\text{-}1}{c_i\ket{\phi_i}};\quad\text{where}
    \sum_{i=0}^{|{\mathcal{H}}|\text{-}1}{c_i^2} = 1;\quad\quad
    c_i \in \mathbb{C}
\end{aligned}
\end{eqnarray*}

The qubit is the quantum analog of a classical bit. It exists in a two-dimensional Hilbert space $ {\mathcal{H}}_2 $, spanned by the basis vectors $ \{ \ket{0}, \ket{1} \} $. Multiple qubits can be combined to form a quantum register (or qregister), whose state resides in the tensor product space $ {\mathcal{H}}_2^{\otimes n} = {\mathcal{H}}_{2^n} $, where $ n $ is the number of qubits. While structurally analogous to classical bit arrays, qubits, and by extension qregisters, can exist in superpositions of all possible states simultaneously. This capability underlies the quantum parallelism that gives quantum computing its potential advantage over classical computation.

To perform computation on a quantum system, it is essential to manipulate the state of a quantum register in a controlled manner. This is done by applying quantum gates, which are unitary operators $U$ that transform an initial state $ \ket{\psi_0} $ into another state $ \ket{\psi_1} $ according to $ U\ket{\psi_0} = \ket{\psi_1}  $. A sequence of quantum gates corresponds to a sequence of unitary operations, applied in succession to the register. If $U_0,U_1,...,U_n$ are such operations, then the final state is given by the following equation; where $U_T$ represents the total transformation applied to the system.

\begin{eqnarray*}
\centering
\begin{aligned}
U_n \cdots U_1 U_0 \ket{\psi_0} = U_T \ket{\psi_0} = \ket{\psi_{n+1}}
\end{aligned}
\end{eqnarray*}

\subsubsection{Random vs. Quantum Walk}
\label{subsubsec:dqw}

A Random Walk (RW) is an iterative stochastic process in which a walker moves randomly through a space by updating its position over time. These spaces are typically discrete sets of positions $ \{x\}$, where ${x \in \Sigma} $ and $ \Sigma $ represent the domain being explored (e.g., $ \Sigma = \mathbb{Z} $ for the one-dimensional case, or $ \Sigma = \mathbb{Z} \times \mathbb{Z} $ for two dimensions). The process consists of two key components: (i) a decider which randomly selects an action (a move) from a set $ \mathcal{A} $ of allowed moves (depending on the topology of $ \Sigma $), and (ii) the current position of the walker, which is updated based on the chosen action. 

\begin{figure}[t]
    \centering
    \includegraphics[width=0.618\linewidth]{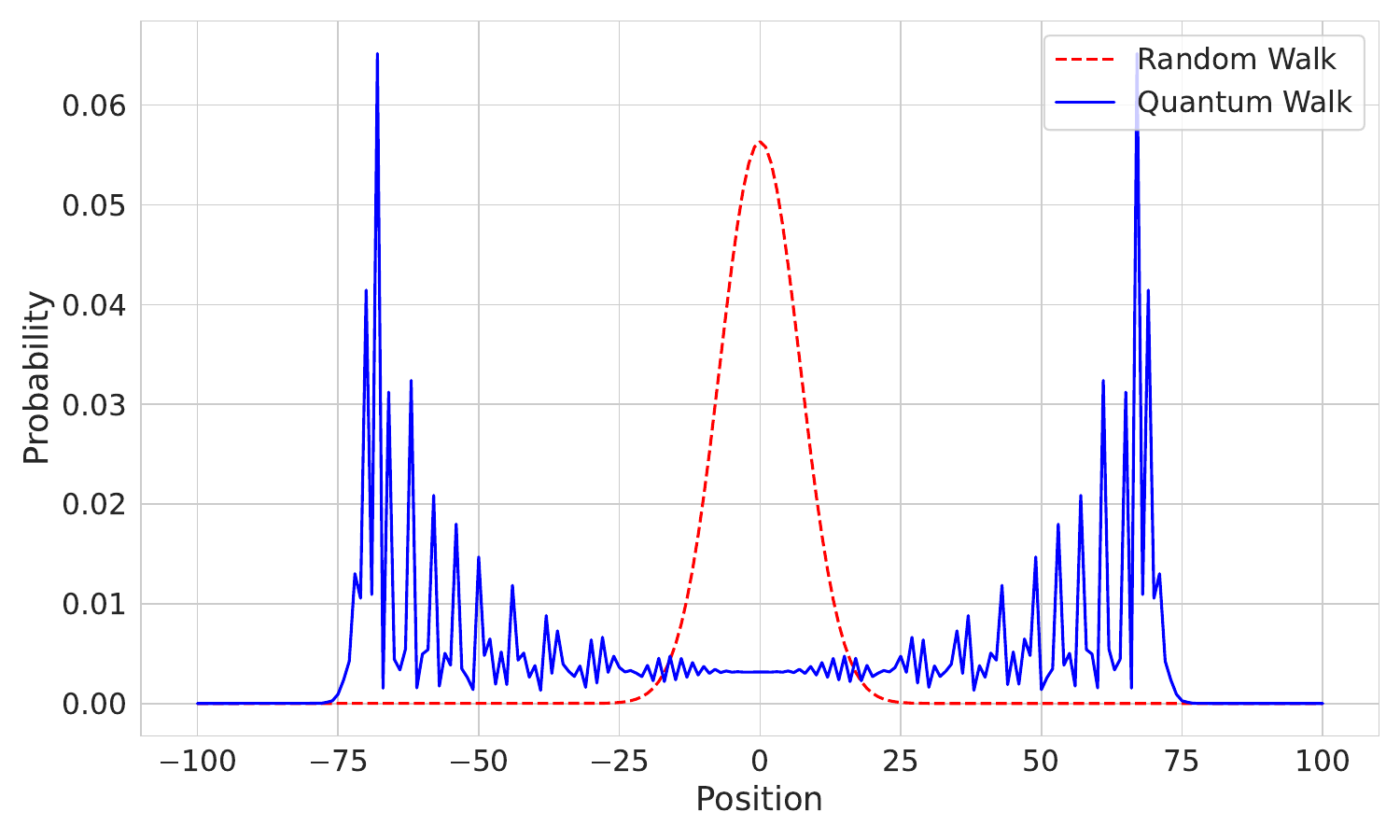}
    \caption{Comparison of the probability distributions over the state space $ \Sigma $ after 100 iterations of a classical Random Walk (red dashed line) and a Discrete Quantum Walk (blue solid line), both starting from the initial position $x_0 = 0$ with equal probability ($ p = 0.5 $) of moving left or right. The quantum version exhibits faster spatial spread and higher variance than its classical counterpart.}
\label{fig:qw_vs_rw}
\end{figure}

After $t$ iterations, the stochastic nature of the process causes the walker’s position to follow a probability distribution over the positions in $ \Sigma $. In the one-dimensional case ($ \Sigma = \mathbb{Z} $), the actions set is defined as $ {\mathcal{A}} =  \{-1, 1\} $, with $ -1 $ representing the leftward movement ($ x \leftarrow x\text{-}1 $) and $ +1 $ the rightward movement ($ x \leftarrow x\text{+}1 $).  Assuming the actions are chosen with equal probability and the initial state is $ x_0 = 0 $, the resulting probability distribution after $t=100$ iterations is shown in Figure \ref{fig:qw_vs_rw} (red dashed line).

The Discrete Quantum Walk (DQW) is the quantum analog of classical Random Walk, and, like its classical counterpart, it consists of two core components: a position register and a decider (or ``coin'') register. These are implemented as quantum registers (qregisters) with associated Hilbert spaces $ {\mathcal{H}}_x = \{\ket{x}\}_{x \in \Sigma} $ for the position and $ {\mathcal{H}}_d = \{\ket{a}\}_{a\in \mathcal{A}} $ for the decider. Together, these registers define the full quantum state of the system. The evolution of the system is driven by a unitary operation $ U $, which applies a single step of the quantum walk. After $ t $ iterations, the system evolves from an initial state $ \ket{\psi_0} $ to a final state $ \ket{\psi_t} = U^t\ket{\psi_0} $.

The operator $U$ can be decomposed into two components: the coin operator $C$, which generates a superposition of possible actions (from the action set $ \mathcal{A} $) in the decider register, and the shift operator $S$, which updates the position register based on the selected action. These components are combined according to \eqref{eq:U_SC}, where $I$ is the identity operator on the position register, and $\otimes$ denotes the tensor product.

\begin{eqnarray}
\label{eq:U_SC}
\centering
\begin{aligned}
    U = S\cdot (C \otimes I)
\end{aligned}
\end{eqnarray}

The coin operator $ C $ plays a central role in the behavior of the Discrete Quantum Walk. In the classical Random Walk, the walker’s position is updated at each step based on a single action selected by the decision maker from the action set $ \mathcal{A} $. In contrast, the quantum version uses the $ C $ operator to place the decider’s register into a superposition of all possible actions in $ \mathcal{A} $. This enables the walker to evolve, in a single iteration, as if it were simultaneously following every available path.

To illustrate the behavior of the QW, we again consider the one-dimensional case, $ \Sigma = \mathbb{Z} $, as previously done for the classical Random Walk. In this setting\footnote{The notation $ H $ and $ T $ originates from the DQW literature, referencing ``Head'' and ``Tail'', the two outcomes of a coin flip, mirroring the role of the decision mechanism.}, we define the decider’s register as $ {\mathcal{H}}_d = \{ \ket{H}, \ket{T} \} $ , where $\ket{H}$ corresponds to a leftward move ($ x \leftarrow x\text{-}1 $) and $ \ket{T} $ to a rightward move ($ x \leftarrow x\text{+}1 $). In this context, the quantum operators $C$ (coin) and $S$ (shift) take the specific forms given in Equation~\eqref{eq:C_S_1D}.

Before analyzing the evolution of the walker's distribution in  $ \Sigma = \mathbb{Z} $, it is helpful to examine the coin operator $ C $. The matrix $ C $ defined in Equation~\eqref{eq:C_S_1D} represents a general coin operator that creates a superposition of the basis states $ \ket{H} $ and $ \ket{T} $, modulated by the parameters $ \theta, \gamma_0 $ and $ \gamma_1 $. The parameter $ \theta $, often referred to as the rotation angle, determines the weighting of the superposition and lies within the interval $ [0, 2\pi) $. For example, when $ \theta = \frac{\pi}{4} $, the matrix reduces to a well-known Hadamard operator, which produces an equal (equiprobable) superposition of the two states. Other values of $ \theta $ introduce a bias, favoring one direction over the other.

The values $ \gamma_0 $ and $ \gamma_1 $, both within the range $ [0, \pi) $, are referred to as the phases of the coin operator. These values influence the global and relative phases of the resulting quantum state, a property that emerges from the fact that quantum states are defined over the complex vector space $ C^n $. Importantly, these phases do not affect the probabilities associated with each basis state in the superposition. As a result, for a given bias (determined by $ \theta $), there exists an infinite family of equivalent coin operations that differ only by their phases.

\begin{eqnarray}
\label{eq:C_S_1D}
\begin{aligned}
    &C = \cos\theta\ket{H}\bra{H} + e^{i\gamma_0}\sin\theta\ket{H}\bra{T} + e^{i\gamma_1}\sin\theta\ket{T}\bra{H} - e^{i(\gamma_0 + \gamma_1)}\cos\theta\ket{T}\bra{T}
    \\\\
    &S = \sum_{\forall x}{\ket{H, x\text{-}1}\bra{H, x}} +
        \sum_{\forall x}{\ket{T, x\text{+}1}\bra{T, x}}
\end{aligned}
\end{eqnarray}

To analyze the evolution of the system in the one-dimensional case $ \Sigma = \mathbb{Z} $, we consider the initial state $ \ket{\psi_0} = \ket{H,0} $ as defined in Equation \eqref{eq:qw_psi0}. Applying the unitary operator $U$ to $ \ket{\psi_0} $ yields the next state $ \ket{\psi_1} $. Recall that the coin operator $C$ creates a superposition over the decider’s basis states, meaning all possible actions are chosen simultaneously. In this case, the walker moves both to the left $ (x \leftarrow x-1) $ and to the right $ (x \leftarrow x+1) $ at once. As shown in Equation~\eqref{eq:qw_psi1}, the resulting state reflects this bidirectional evolution, with the amplitude (and thus probability) of each position determined by the action's contribution in the coin superposition.

Reapplying the operation $U$ to $ \ket{\psi_1} $, the system further evolves from each of its current positions ($ \ket{H, -1} $ and $ \ket{H,+1} $), resulting in the state $ \ket{\psi_2} $ given in Equation \eqref{eq:qw_psi2}. Now, the probability of measuring each state corresponds to the combined probability of each move to reach that position; as reflected by the probability amplitude of the state $ \ket{H, -2} $, the result of two moves to the left with probability amplitude $ \cos \theta $. However, it is important to note how, depending on the state of the decider register, the probability of a move can vary. This effect can be noticed in the state $ \ket{T, +2} $ where, despite being a result of two movements to the right, both movements have different probability amplitudes; $ \cos \theta $ and $ e^{i(\gamma_0 + 2\gamma_1)} \sin \theta $ respectively. This, combined with the negative terms generated by the $ C $ operator in the probability amplitudes, can generate constructive and destructive interferences in following iterations.

\begin{eqnarray}
\begin{aligned}
    \ket{\psi_0} = \ket{H}\ket{0} = \ket{H, 0}
\end{aligned}
\label{eq:qw_psi0}
\end{eqnarray}

\begin{eqnarray}
\begin{aligned}
    \ket{\psi_1} = U\ket{\psi_0} = \cos\theta\ket{H,-1} + e^{i\gamma_1}\sin\theta\ket{T,+1}
\end{aligned}
\label{eq:qw_psi1}
\end{eqnarray}

\begin{eqnarray}
\begin{aligned}
    \ket{\psi_2} = &\cos^2\theta\ket{H,-2} + e^{i(\gamma_0+\gamma_1)}\sin^2\theta\ket{H,0} +\\
    &e^{i\gamma_1}\sin\theta\cos\theta\ket{T,0} - e^{i(\gamma_0+2\gamma_1)}\sin\theta\cos\theta\ket{T,+2}
\end{aligned}
\label{eq:qw_psi2}
\end{eqnarray}

Figure~\ref{fig:qw_vs_rw} compares the probability distribution of the walker in a classical Random Walk (RW, red dashed line) and a Quantum Walk (QW, blue solid line), both evaluated after $ t=100 $ iterations from the same initial position $ x_0 = 0 $. As observed, the QW explores a significantly broader region of the space $ \Sigma $, reaching positions much farther from the origin compared to the RW under the same number of steps. This reflects the greater dispersion achieved by the quantum process. Analytically, the dispersion (standard deviation) of a classical RW scales as $ {\mathcal{O}}(\sqrt{t}) $ whereas for a QW it scales as $ {\mathcal{O}}(t)$, demonstrating a quadratic speedup in spatial exploration. This acceleration arises from the QW's ability to leverage quantum superposition: the decider register evolves into a superposition of actions, enabling the walker to propagate in multiple directions simultaneously. This behavior is clearly illustrated in the first two steps of the system’s evolution (see Equations~\eqref{eq:qw_psi1} and \eqref{eq:qw_psi2}).

\section{From Quantum Walk to Quantum MCMC}
\label{sec:proposal}

The goal of this work is to design a quantum version of the Markov Chain Monte Carlo (MCMC) algorithm based on the DQW algorithm. To achieve this, it is necessary to introduce into the QW circuit the acceptance probability $ A $ (Subsection \ref{subsec:mcmc}) in such a way that it biases the decision to accept or reject a move. We have achieved this by adding a single qubit register, referred to as the “coin”, and which is used as a control qubit over the $ S $ operator.

A controlled operator, or $ c\text{-}U $ for short, allows to control the application of an $ U $ operator on other registers depending on the state of a qubit (or qubits) acting as a controller. By convention, the state $ \ket{1} $ represents the application of operator $ U $ and $ \ket{0} $ the non-application. The interesting thing about these operators occurs when working with control qubits in superposition; resulting in a circuit state combination of having applied $ U $ and not having applied it. As can be seen in Equation \eqref{eq:cU_effect}, the mix of these two options comes as a function of the superposition in the control qubit; that is, in the probability $ p $ of obtaining a $ \ket{1} $ in the control qubit.

\begin{eqnarray}
\label{eq:cU_effect}
\begin{aligned}
    c\text{-}U\ket{x}\left(\sqrt{1\text{-}p}\ket{0} + \sqrt{p}\ket{1}\right) =
    \left( \sum_{x}{\ket{x}} \bra{x} \right) 
    \otimes \sqrt{1\text{-}p}\ket{0}\bra{0} \\ +
    \left( \sum_{x}{U\ket{x} \bra{x}}\right) 
    \otimes \sqrt{p}\ket{1}\bra{1}
\end{aligned}
\end{eqnarray}

With this mechanism, and by regulating the probability $ p $ in the new register $ \ket{\text{coin}} $ with the information of the current position of the circuit, we can implement the probability of acceptance of the MCMC methods to make the system  behaves like an MCMC process, generating a user-defined target distribution $ f $, which can then be sampled through multiple measurements of the position register, all while preserving the enhanced exploration capabilities provided by the quantum walk.

\subsection{Proposed circuit}

As outlined in Subsection \ref{subsec:mcmc}, every MCMC process consists of three steps: generating a movement proposal, calculating the probability of accepting the proposed move, and updating the current position if the proposal is accepted. The quantum circuit we propose, shown in Figure \ref{fig:proposed_circ}, is designed to emulate this behavior through four core quantum gates:  (i) TRIAL, (ii) DISC, (iii) C GROUP and (iv) SHIFT. Before describing the function of each gate, it is important to clarify the idea behind the design and the specific role of each quantum register involved in the circuit.

The proposed circuit consists of the following quantum registers:

\begin{itemize}
    \item The action register $ \ket{a} $ represents an action from the set $ \mathcal{A} $ derived from the state space $ \Sigma $. Due to quantum superposition, this register can simultaneously encode a distribution over all possible actions, enabling efficient parallel exploration.

    \item The position register $ \ket{x} $ stores the current position of the walker in the space $ \Sigma $. Similar to $ \ket{a} $, it benefits from quantum superposition to represent multiple positions simultaneously.

    \item The trial register $ \ket{t} $ is an auxiliary register used to store the trial state proposed based on the current values of $ \ket{a} $ and $ \ket{x} $.

    \item The acceptance register $ \ket{\text{acc}} $ encodes the discretized acceptance probability associated with the proposed move. Its value is computed using the DISC gate.

    \item The coin register $ \ket{\text{coin}} $ acts as a quantum decision mechanism, determining whether the proposal will be accepted or rejected, using the probability encoded in $ \ket{\text{acc}} $, to modulate the final outcome.
\end{itemize}

At the start of the circuit's execution, all the quantum registers are initialized to the state $ \ket{0} $, except for the position register $ \ket{x} $, which may be initialized to any arbitrary value representing the initial position of the walker. In this case, to avoid the initial position selection problem encountered in the classical MCMC, the register $ \ket{x} $ is initialized in an equiprobable superposition of all its states, effectively considering every state as a possible starting point.

\begin{figure*}[htbp]
    \centering
    \scalebox{0.66} {
    \begin{quantikz}
        \ket{a} & \gate{H} & \gate[3]{\text{TRIAL}} &&&&&&& \gate[3]{\text{TRIAL}^\dagger} & \gate[2]{\text{SHIFT}} &\gate{\text{SWAP}}&\\
        \ket{x} &&& \gate[3]{\text{DISC}} &&&&& \gate[3]{\text{DISC}^\dagger} &&&&\\
        \ket{t} &&&&&&&&&&&&\\
        \ket{\text{acc}} &&&& \octrl{1}{\text{0}}\gategroup[2, steps=4]{C GROUP} & \octrl{1}{\text{1}} & \octrl{1}{\text{2}} & \octrl{1}{\text{3}} &&&&&\\
        \ket{\text{coin}} &&&& \gate{C_0} & \gate{C_1} & \gate{C_2} & \gate{X} &&& \ctrl{-3} & \gate{\text{SWAP}}&
    \end{quantikz}
    }
    \vspace{0.3cm}
    \caption{Quantum circuit for the proposed QMCMC algorithm. The circuit is composed of four main operations: TRIAL, DISC, C GROUP (a collection of controlled C gates) and SHIFT. Gates marked with the symbol  $ \dagger $ represent the inverses of their respective operations and are included to enable the reversibility and iterability of the process.}
\label{fig:proposed_circ}
\end{figure*}

The algorithm begins by applying to the register $ \ket{a} $ an operator that sets a superposition state over all possible actions. Depending on the operator applied, the superposition in $ \ket{a} $ can acquire different forms, biasing the probability of the different directions of the space. In this sense, the distribution in $ \ket{a} $ can be understood as the generating function $ g $ mentioned previously in Subsection \ref{subsec:mcmc}. This step is analogous to the original QW, where we apply the $ C $ operator to create a superposition on the decider register to value multiple directions at the same time. In this paper, we assume a scenario where we do not know anything about the target distribution, only the function $ f $ that defines it. In this scenario, there is no reason to bias in favor of the probability of any of the directions of the space, so the best strategy is to equiprobably explore all directions. To do so, we apply a Hadamard gate (denoted by $ H $) to generate the superposition representing such a scenario.

In the next step, as in the MCMC algorithm, a new trial is generated. This operation is performed using the TRIAL gate, which takes as input the registers $ \ket{a} $ and $ \ket{x} $, and places the resulting proposal in the output  register $ \ket{t} $. Considering a discrete space $ \Sigma = \mathbb{Z} $, and actions defined by $ {\mathcal{A}} = \{ \ket{0}, \ket{1} \} $, where $ \ket{0} $ represents the movement to the left and $ \ket{1} $ the movement to the right, based on the values of $ \ket{a} $ and $ \ket{x} $ registers, the trial value can be expressed by $ x\text{-}(\text{-}1)^a $. Therefore, the operation TRIAL takes the form of Equation \ref{eq:trial_gate}. The last sum in the value of the register $ \ket{t} $ is necessary to preserve the unitarity of the operation.

\begin{eqnarray}
\label{eq:trial_gate}
\begin{aligned}
    \text{TRIAL} = 
    \sum_{a, x, t}
    {\ket{a, x, x\text{-}(\text{-}1)^a\text{+}t}}
    \bra{a, x, t}
\end{aligned}
\end{eqnarray}

The second step involves calculating the acceptance probability $ A $ and mapping it onto the quantum register $ \ket{\text{acc}} $. Remember that because we applied a Hadamard gate to the register $ \ket{a} $, we are considering a generating distribution $ g $ that is uniform over all possible actions; that is, a kind of symmetrical distribution. For this reason, and as mentioned in Subsection \ref{subsec:mcmc}, Equation \eqref{eq:acc_f} can be simplified to the one shown in Equation \eqref{eq:disc_gate}. Since quantum registers cannot represent continuous values directly, it is necessary to discretize the range $(0,1]$ into multiple finite intervals, where each interval has a unique index associated with it. The function that determines the size of each interval and the index that represents it is what we call the discretization rule $ D $. This mapping of the acceptance probability is carried out with the DISC operation; Equation \eqref{eq:disc_gate}, assigning in $ \ket{\text{acc}} $ the corresponding interval index based on the value of $ \ket{x} $ and $ \ket{t} $. Note that the number of intervals that we can represent is determined by the number of possible states of the register $ \ket{\text{acc}} $, so to obtain a good approximation of the acceptance probability a sufficiently large $\ket{\text{acc}} $ register is needed.

\begin{eqnarray}
\label{eq:disc_gate}
\begin{aligned}
\text{DISC} = \sum_{x, t, \text{acc}}
    {\ket{x, t, D(x, t)\text{+acc}}}
    \bra{x, t, \text{acc}};\quad\quad
&D(x,t) = \left\{
    \begin{array}{ll}
    0, & A(t, x) < 0.33 \\
    1, & 0.33 \leq A(t, x) < 0.66 \\
    2, & 0.66 \leq A(t, x) < 1 \\
    3, & A(t, x) \geq 1
    \end{array}\right.\\\\
&A(t, x) = \min\left(1, \dfrac{f(t)}{f(x)}\right)
\end{aligned}
\end{eqnarray}

Once the acceptance probability has been discretized, the quantum register $ \ket{\text{acc}} $ is used as a control to activate the corresponding $ C_i $ gate on the quantum register $ \ket{\text{coin}} $. This operation creates a superposition in $ \ket{\text{coin}} $ that represents the acceptance probability associated with movement from $ \ket{x} $ to $ \ket{t} $ according to the acceptance probability $ A $ of Equation \eqref{eq:disc_gate}. In the implementation, the angles $ \theta_i $ are preset to match the average value of each discretized interval, aiming to minimize rounding errors. It is important to highlight that, in general, the last interval is used to represent the acceptance of the trial ($A(t,x) \geq 1 $), so a NOT gate only is needed (X gate in quantum notation), as it only needs to flip the state of $ \ket{\text{coin}} $ to $ \ket{1} $.

Finally, once the quantum register $ \ket{\text{coin}} $ encodes the probability of acceptance of the trial, the system updates its position within the state space. This update is controlled by the SHIFT gate, which operates on the registers $ \ket{a} $, $ \ket{x} $, and $ \ket{\text{coin}} $. Specifically, the gate uses the value in $\ket{\text{coin}}$ to determine whether to perform the shift: if the proposal is accepted ($\ket{\text{coin}}$ = $\ket{1}$), the register $\ket{x}$ is updated to reflect the new state; otherwise, it remains unchanged. The mathematical definition of the SHIFT gate for the one dimensional case is given in Equation \eqref{eq:shift_gate}. The new state in $ \ket{x} $ is the same explained in the TRIAL operation, but now applied on $ \ket{x} $ instead of on $ \ket{t} $.

\begin{eqnarray}
\label{eq:shift_gate}
\begin{aligned}
    \text{SHIFT} =
    \left( \sum_{a, x}{\ket{a, x}} \bra{a,x} \right) 
    \otimes \ket{0}\bra{0} +
    \left( \sum_{a, x} {\ket{a, x-(-1)^a} \bra{a,x}}\right) 
    \otimes \ket{1}\bra{1}
\end{aligned}
\end{eqnarray}

A key component of iterative quantum algorithms is the process known as uncomputation, which allows previously calculated values to be reversed and reset. This is essential for reusing quantum registers without disturbing the superposition states that may be entangled with other parts of the system. In the proposed circuit, uncomputation is applied to the registers $ \ket{t} $ and $ \ket{\text{acc}} $, returning them to the $ \ket{0} $ state. Since all information required for the next iteration is encoded in the $ \ket{\text{coin}} $ register, the inverse gates $ \text{TRIAL}^\dagger $ and $ \text{DISC}^\dagger $ are used to safely undo the previous operations.

However, the registers $ \ket{a} $ and $ \ket{\text{coin}} $ cannot be recomputed without disrupting the quantum state of $ \ket{x} $, which now stores the computed probability distribution. Therefore, to support multiple iterations, these registers must be replaced with new ones initialized to $ \ket{0} $. This is done using SWAP gates, which exchange the entangled registers with clean ones prepared for reuse. The reason for not using the same register $ \ket{a} $ as in the original QW circuit is to avoid an observable effect in Equation \eqref{eq:qw_psi2}. In this equation it can be seen how for the state $ \ket{T, +2} $, a consequence of moving twice to the right in the example, it assigns a different probability to both movements. This phenomenon is not desirable and is due to the unitary nature of the operators in quantum computing. Replacing the qubits of the register $ \ket{a} $; without altering the superposition of the old qubits, allows to increase the state space of the circuit and assign the correct probability to each movement at any given time.

One limitation of this approach is that the total number of qubits increases linearly with the number of iterations following Equation \eqref{eq:qubit_increase}, where $ t $ is the number of iterations and $ n_a $, $n_x$ and $n_{\text{acc}}$ the sizes of each qregister of the circuit\footnote{Since the size of qregister $ \ket{t} $ is always $ n_t = n_x $, the size of qregister $ \ket{x} $ is considered twice in the expression above.}. In Section~\ref{sec:proposal} we have defined the circuit resulting of considering the one dimensional case when $ n_a = 1 $. But, in the general case when considering a $ d $-dimensional space, to represent all the moves it is necessary that the qregister $ \ket{a} $ has, at least, a size of $ n_a = \lceil\log_2(2d\cdot l)\rceil $; where $ d $ is the dimensionality of the space (each dimension has two directions: towards the negatives and towards the positives), and $ l $ is the number of moves per direction.

\begin{eqnarray}\label{eq:qubit_increase}
    (n_a+1)t+2n_x+n_{\text{acc}}
\end{eqnarray}

\section{Results}
\label{sec:results}

Before presenting and analyzing the results obtained on simulation, three important points should be considered. First, the label that identifies each quantum state in the circuit does not directly correspond to its actual real-valued representation. In the experiments, the quantum state $ \ket{x} = \ket{0} $ maps to the real value -5, and the state $ \ket{x} = \ket{|{\mathcal{H}}_x| - 1} $ maps to 5. The intermediate values are derived via linear interpolation.

Second, note that the target function $ f $ is a continuous probability density function defined over an unbounded or very large domain. In order to represent it using a quantum register, the domain must be truncated to a finite interval, for example, $[-5,5]$, and discretized into evenly spaced segments, each corresponding to a basis state of the quantum register $ \ket{x} $. As a result, each state encodes the probability mass within a fixed-width subinterval. The discretized version of the target distribution, limited to the chosen interval, is referred to as the expected distribution, and it is this shape that the circuit aims to approximate. Importantly, while the resulting probability distribution always sums to 1, it only captures the structure of $ f $ within the truncated range.

Finally, the circuit has been iterated until the resulting distribution codified in the position register $ \ket{x} $ superposition converged. 

In this context, we ascertain convergence when the distance between the distributions of consecutive iterations is below certain value $ \epsilon $, where we use the total variation distance to measure the difference between two distributions $ d_1, d_2 : \sum_{x}{|d_1(x) - d_2(x)|} $. In each case, the value of $ t $ where the distribution achieves convergence is used to evaluate the convergence rate of the circuit.

\subsection{Analysis of the resulting distribution}
\label{subsec:distro_analysis}

\begin{figure*}[htbp]
\centering
\subfloat[$ \mathcal{N}(0; 1);\quad |{\mathcal{H}}_x|  = 2^4 $]{
    \label{fig:conver_g0_1_4}
    \includegraphics[width=0.32\linewidth]{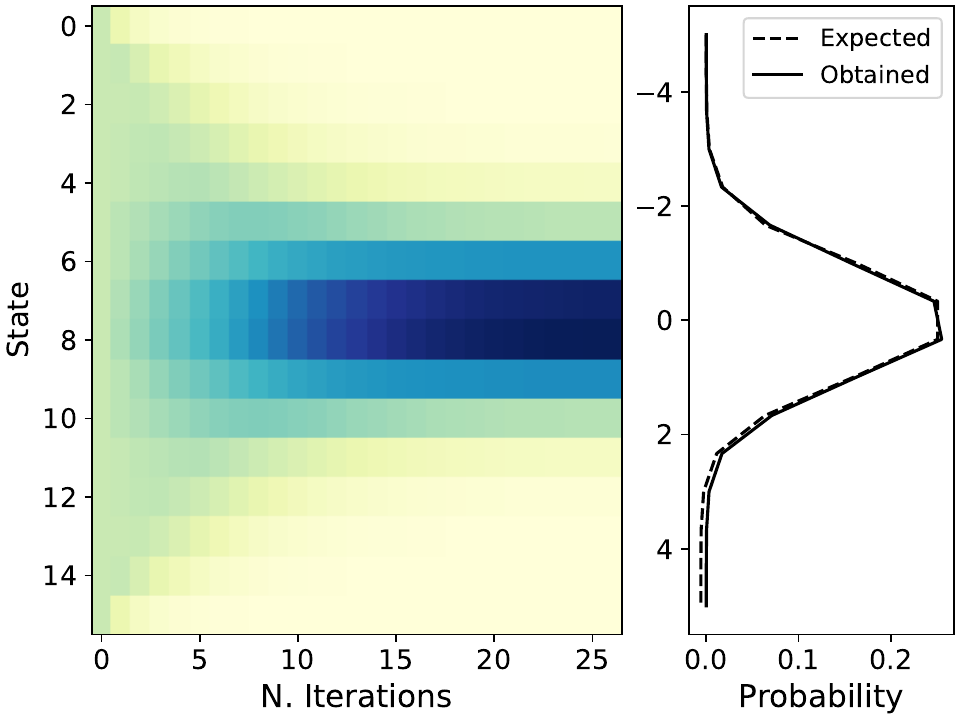}
}
\subfloat[$ \mathcal{N}(2; 1);\quad |{\mathcal{H}}_x|  = 2^4 $]{
    \label{fig:conver_g2_1_4}
    \includegraphics[width=0.32\linewidth]{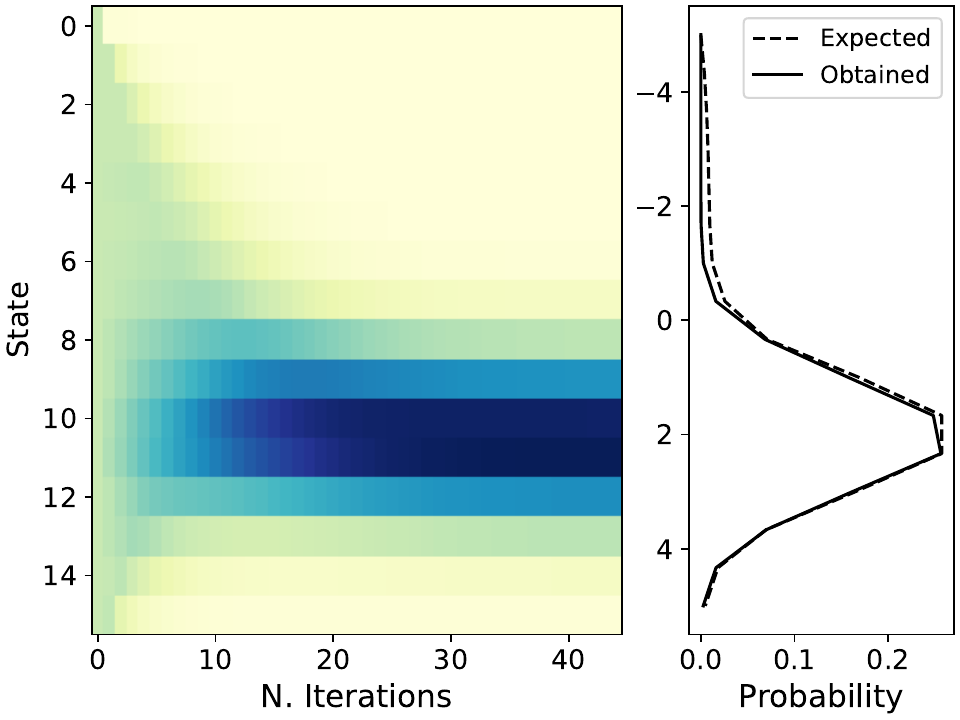}
}
\subfloat[$ \mathcal{N}(0; 1)+\mathcal{N}(3; 0.5);\quad |{\mathcal{H}}_x|  = 2^5 $]{
    \label{fig:conver_g01_35_5}
    \includegraphics[width=0.32\linewidth]{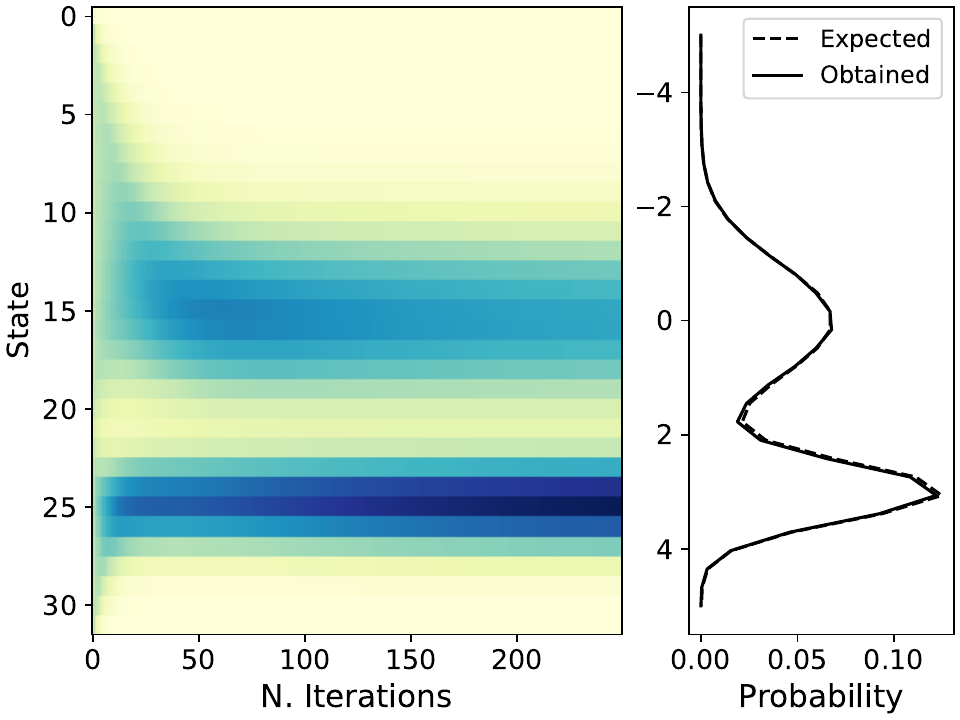}
}

\subfloat[$ \mathcal{N}(0; 1);\quad |{\mathcal{H}}_x|  = 2^5 $]{
    \label{fig:conver_g0_1_5}
    \includegraphics[width=0.32\linewidth]{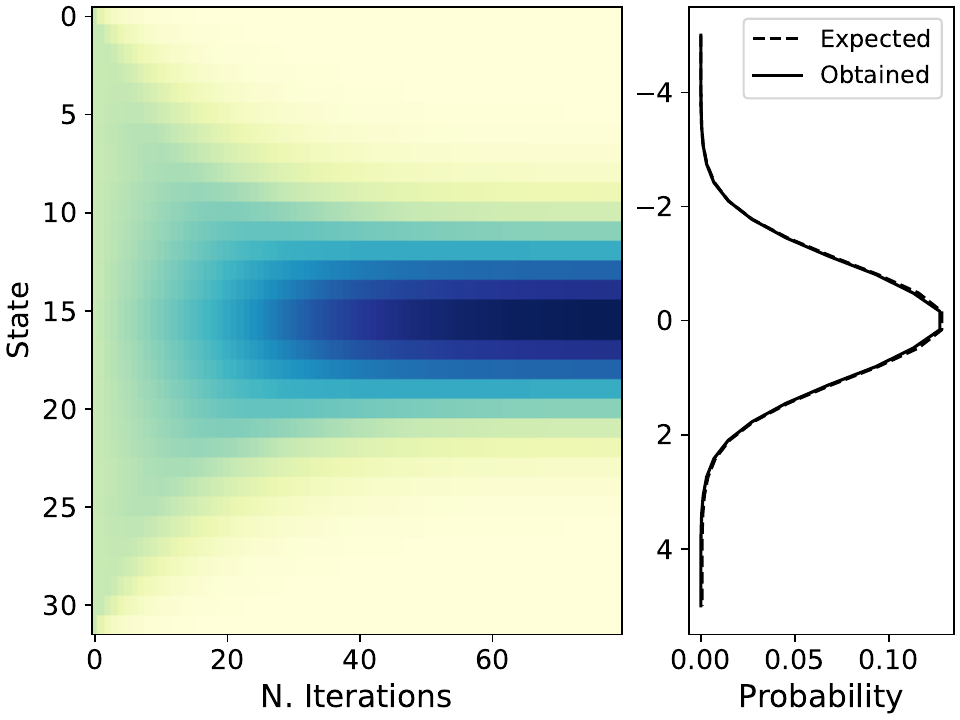}
}
\subfloat[$ \mathcal{N}(2; 1);\quad |{\mathcal{H}}_x|  = 2^5 $]{
    \label{fig:conver_g2_1_5}
    \includegraphics[width=0.32\linewidth]{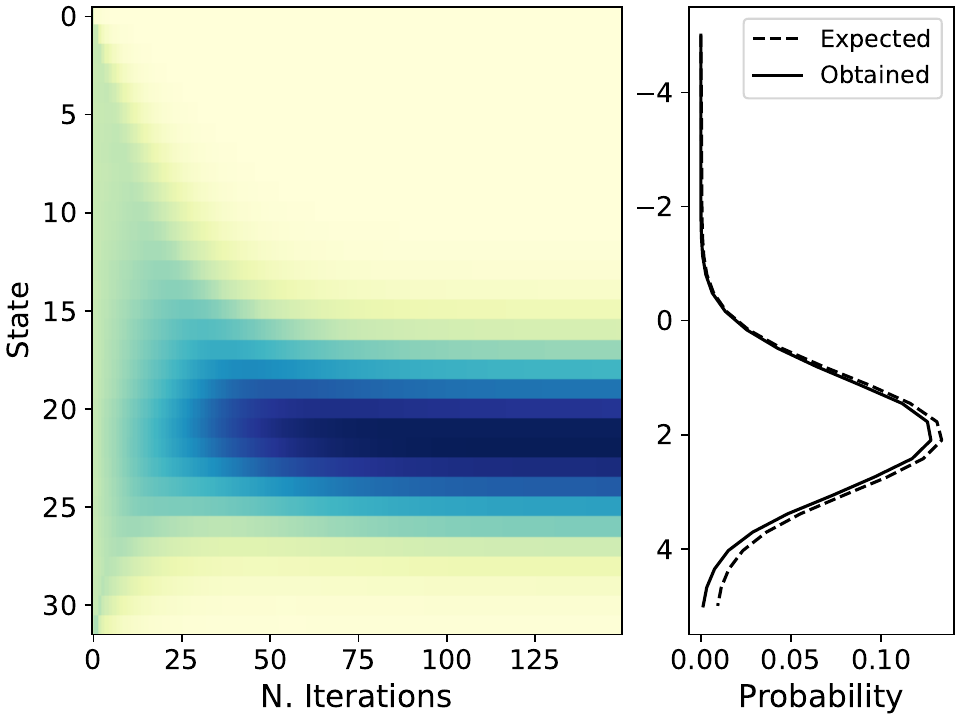}
}
\subfloat[$ \mathcal{N}(0; 1)+\mathcal{N}(3; 0.5);\quad |{\mathcal{H}}_x|  = 2^7 $]{
    \label{fig:conver_g01_35_7}
    \includegraphics[width=0.32\linewidth]{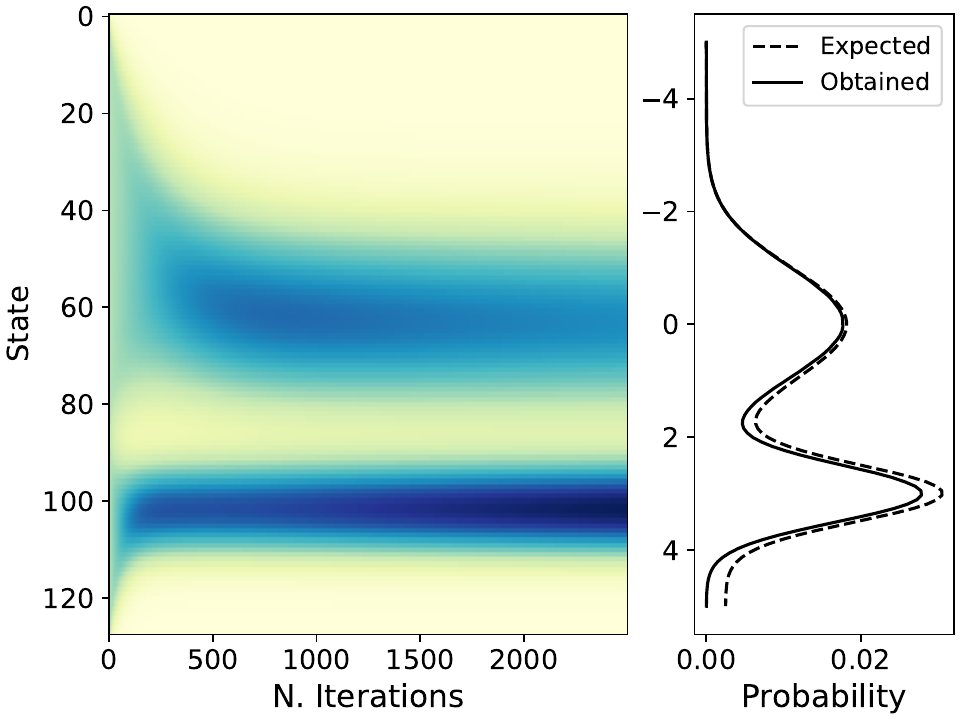}
}
\vspace{0.2cm}
\caption{Simulation results illustrating the circuit's ability to approximate various target distributions. Each subplot contains: a heatmap representing the evolution of the quantum state's probability distribution over successive iterations (columns), where yellow indicates low probability and blue indicates high probability and a plot comparing the expected discretized distribution (dashed line) with the obtained distribution from the quantum circuit (solid line). The function being approximated and the size of the position register $ \ket{x} $ are specified in each subplot caption. It can be observed that the circuit successfully approximates the shape of the target function $ f $, which in these experiments corresponds to individual and mixtures of gaussians. Increasing the size of the register $ \ket{x} $ enhances the resolution of the approximation, although it also leads to a higher number of iterations required to reach convergence.}

\label{fig:conver_figs}
\end{figure*}

Figure \ref{fig:conver_figs} presents six experiments performed using different parameters: (i) different target functions $f$, including single Gaussian distributions $ \mathcal{N}(\mu,\sigma) $ and mixtures of Gaussians, and (ii) varying sizes of the quantum register $ \ket{x} $. In all cases, the size of the register $ \ket{\text{acc}} $ was set equal to the size of $ \ket{x} $, that is, $ |{\mathcal{H}}_\text{acc}| = |{{\mathcal{H}}_x}| $. In Subfigure \ref{fig:conver_g0_1_4}, the circuit approximates $ \mathcal{N}(0,1) $ and the resulting distribution closely matches the expected one in both the mean $ \mu $ and the standard deviation $ \sigma $. A similar result is seen in Subfigure \ref{fig:conver_g2_1_4} for $ \mathcal{N}(2;1) $, where the distribution is shifted but also generates a good approximation. These results show that the circuit captures the structure of the target function $f$ within the interval. However, due to the limited representational capacity of $ \ket{x} $, the discretization of $ f $ becomes noticeable.

Increasing the size of the register $ \ket{x} $, see Subfigures \ref{fig:conver_g0_1_5} and \ref{fig:conver_g2_1_5}, results in a more accurate approximation of the continuous function $ f $. However, this improvement comes at the cost of a higher number of iterations required for convergence. This observation suggests a direct relationship between the size of $ \ket{x} $ and the convergence speed: larger registers enable finer discretization but slow down the convergence.

In addition to the size of the register $ \ket{x} $, the complexity of the function $ f $ is also a key factor to determine the number of iterations to reach convergence. Subfigures \ref{fig:conver_g01_35_5} and \ref{fig:conver_g01_35_7}, illustrating a mixture of Gaussians $ \mathcal{N}(0;1) + \mathcal{N}(3;0.5) $, show a notable increase in the number of iterations needed. The last simulation (Subfigure \ref{fig:conver_g01_35_7}) is particularly insightful for two reasons: (i) it clearly exposes the significant impact of the complexity of $ f $ on convergence, and (ii) it reveals the circuit’s decreasing convergence rate in the final iterations. While the initial evolution quickly approximates the expected distribution, progress slows considerably as the approximation improves.

In general, the simulations confirm that the circuit successfully approximates the target distribution $ f $ effectively. However, the convergence rate remains a limiting factor. In all cases, the number of iterations required to achieve convergence far exceeds the number of states represented in $ \ket{x} $, which suggests potential limitations in terms of scalability. Applying the circuit to more realistic problems with large and complex state spaces could be challenging, as it may require a significantly higher number of iterations.

\subsection{Effects of the qregister \texorpdfstring{$ \ket{\text{acc}} $}{acc}}
\label{subsec:discret_effects}

\begin{figure*}[t]
\centering
\subfloat[$ |{\mathcal{H}}_\text{acc}| = 2^2 $;\quad (4 intervals)]{
    \label{fig:discret_4}
    \includegraphics[width=0.32\linewidth]{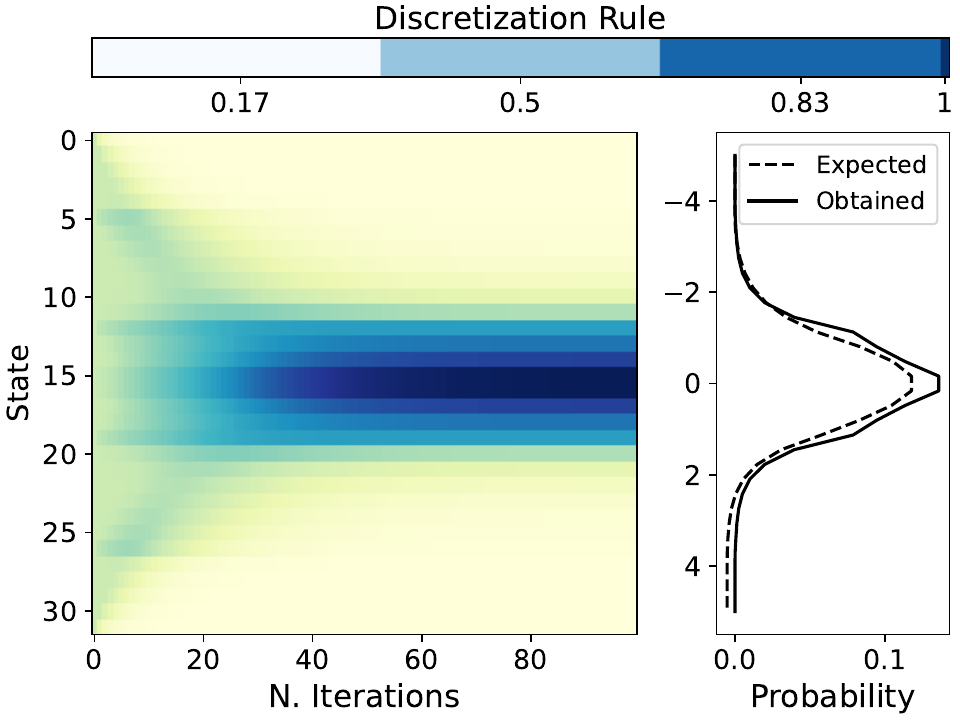}
}
\subfloat[$ |{\mathcal{H}}_\text{acc}| = 2^3 $;\quad (8 intervals)]{
    \label{fig:discret_8}
    \includegraphics[width=0.32\linewidth]{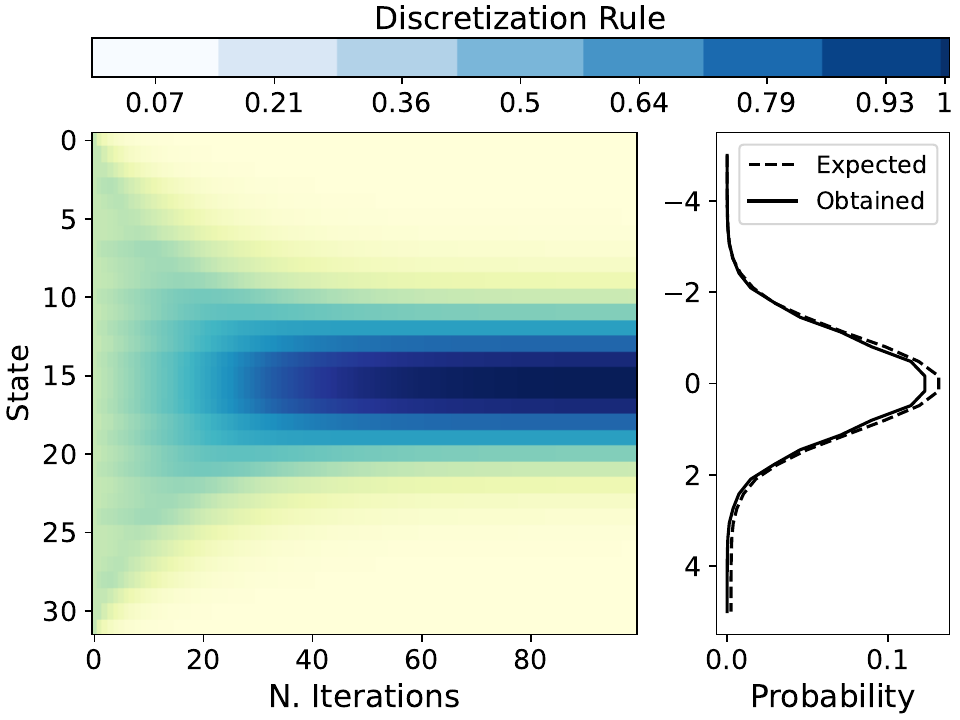}
}
\subfloat[Continous case]{
    \label{fig:discret_none}
    \includegraphics[width=0.32\linewidth]{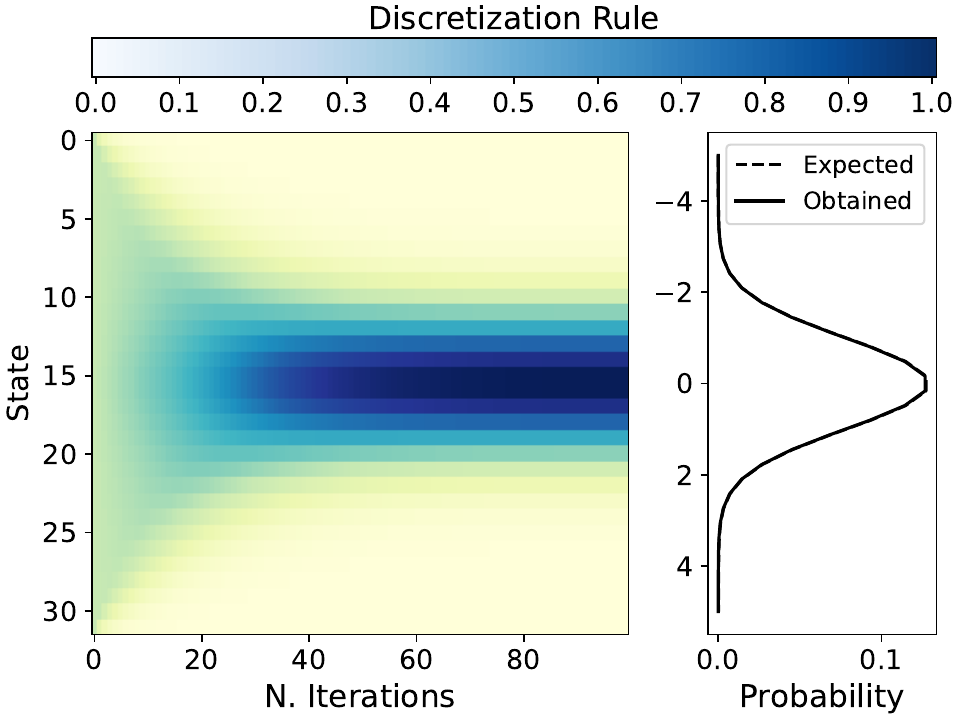}
}
\vspace{0.2cm}
\caption{Impact of the size of the $ \ket{\text{acc}} $ register (which encodes the acceptance probability) on the circuit's ability to approximate the target distribution $ \mathcal{N}(0; 1)$. Subfigure (a) uses a coarse discretization (4 intervals), (b) uses 8 intervals, and (c) presents a theoretical non-discretized case. The plots show how increasing the resolution of acceptance probability improves fidelity to the expected distribution.}

\label{fig:discret_figs}
\end{figure*}

Another adjustable parameter in the circuit is the size of the quantum register $ \ket{\text{acc}} $, which controls the discretization level of the acceptance probability. Experiments conducted with varying $ \ket{\text{acc}} $ register sizes have shown that increasing this size improves the level of detail in the circuit's approximation of the target distribution $ f $. The results of these simulations are shown in Figure \ref{fig:discret_figs}.

Subfigure \ref{fig:discret_none} presents the theoretical scenario in which the acceptance probability of the coin is not discretized. Although such operation cannot be implemented using unitary quantum gates in practice, simulation allows us to explore this idealized case. As expected, the resulting distribution closely follows a smooth form of the target $ f $. When the discretization is introduced to make the gate unitary, some detail is lost, as shown in Subfigure \ref{fig:discret_4}. However, as with the quantum register $ \ket{x} $, increasing the size of $ \ket{\text{acc}} $ improves the level of detail of the result (see Subfigure \ref{fig:discret_8}). Furthermore, additional experiments confirm that larger register sizes consistently lead to better approximations of the target distribution.

All simulations in this section assume the discretization rule $ D $ defined in Equation \eqref{eq:disc_gate}, which uniformly divides the range $ [0, 1) $ into equally sized, symmetrically distributed intervals. However, the method is not limited to this approach, alternative discretization schemes can be applied. Experiments conducted with different discretization strategies showed no significant variation in the convergence speed or the quality of the resulting distributions.

The complexity of the function $ f $ plays an important role when choosing the size of the register $ \ket{\text{acc}} $. Functions that exhibit high multimodality or abrupt trend changes require finer discretization, i.e., more intervals, to capture their structure. As a rule of thumb, we found that setting the size of $ \ket{\text{acc}} $ equal to that of $ \ket{x} $ typically provides sufficient resolution for accurate representation. However, for simpler distributions, such as those shown in Subfigure \ref{fig:discret_8}, a smaller quantum register may still obtain good approximation results.

\subsection{Effect of the size of $ \ket{a} $ on convergence speed}
\label{subsec:converg_improve}

As highlighted at the end of Subsection \ref{subsec:distro_analysis}, and shown in Subfigure \ref{fig:conver_g01_35_7}, the circuit exhibits a slow convergence rate, requiring a number of iterations $ t $ significantly greater than the number of representable states in the register $ \ket{x} $. This effect is particularly noticeable in that case, but similar behavior is observed across all evaluated configurations. This limitation reduces the applicability of the circuit in real-world scenarios, where the state space is typically large. Therefore, reducing the number of iterations—that is, increasing the convergence speed—is crucial to improving the usability of the proposed method. To address this, we modify the register  $ \ket{a} $ to allow for a larger set of possible moves within a single iteration.

As described in Section~\ref{sec:proposal}, the original circuit allows each position to move only to its immediately adjacent states. This constraint mirrors the behavior of classical MCMC algorithms, where each proposed move requires evaluating the function $ f $, typically at a high computational cost. However, in the quantum setting, this limitation can be relaxed: superposition enables simultaneous evaluation of $ f $ at multiple positions, as already exploited through the register $ \ket{a} $. By increasing the size of $ \ket{a} $, we can encode a wider set of candidate moves and explore the state space more broadly in each iteration.

With this modification, the circuit is capable of evaluating up to $|\mathcal{H}_a| = |\mathcal{H}_x| / 2$ proposals per iteration. Equation \eqref{eq:new_trial_shift} defines the updated TRIAL and SHIFT gates that implement this extended behavior; where the indicator function $ \mathbf{1}_{\{a \geq \frac{|\mathcal{H}_a|}{2}\}} $ (a binary value that takes value 1 if the condition is fulfilled) avoids movement from any $ \ket{x} $ to itself. Ideally, one would allow unrestricted movement across the entire state space—i.e., setting $\mathcal{H}_a = \mathcal{H}_x$. However, we have not identified a unitary operation that achieves this, at least within the architectural constraints of the circuit shown in Figure \ref{fig:proposed_circ}.

\begin{eqnarray}
\label{eq:new_trial_shift}
\begin{aligned}
    \text{TRIAL} &= 
    \sum_{a, x, t}
    {\ket{a, x, x+a+\mathbf{1}_{\{a \geq \frac{|\mathcal{H}_a|}{2}\}}-\frac{|\mathcal{H}_a|}{2}+t}}
    \bra{a, x, t}\\\\
    \text{SHIFT} =
    &\left( \sum_{a, x}{\ket{a, x}} \bra{a,x} \right) 
    \otimes \ket{0}\bra{0} +\\
    &\left( \sum_{a, x} {\ket{a, x+a+\mathbf{1}_{\{a \geq \frac{|\mathcal{H}_a|}{2}\}}-\frac{|\mathcal{H}_a|}{2}} \bra{a,x}}\right) 
    \otimes \ket{1}\bra{1}
\end{aligned}
\end{eqnarray}

\begin{figure*}[t]
\centering
\subfloat[$ |\mathcal{H}_a| = 2^1 $]{
    \label{fig:improve_1}
    \includegraphics[width=0.32\linewidth]{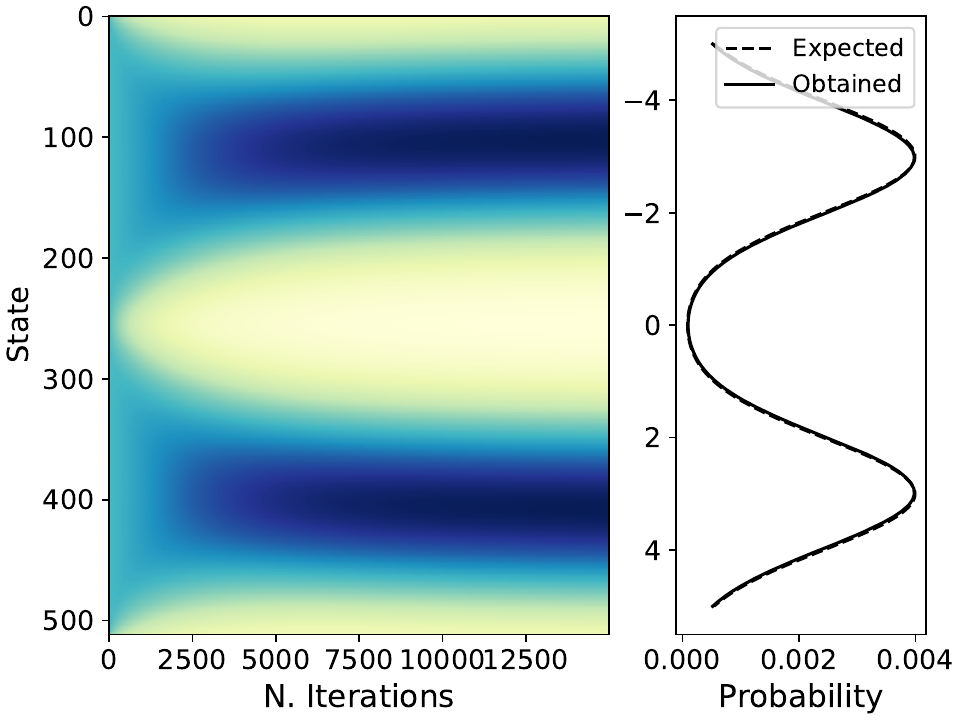}
}
\subfloat[$ |\mathcal{H}_a| = 2^2 $]{
    \label{fig:improve_2}
    \includegraphics[width=0.32\linewidth]{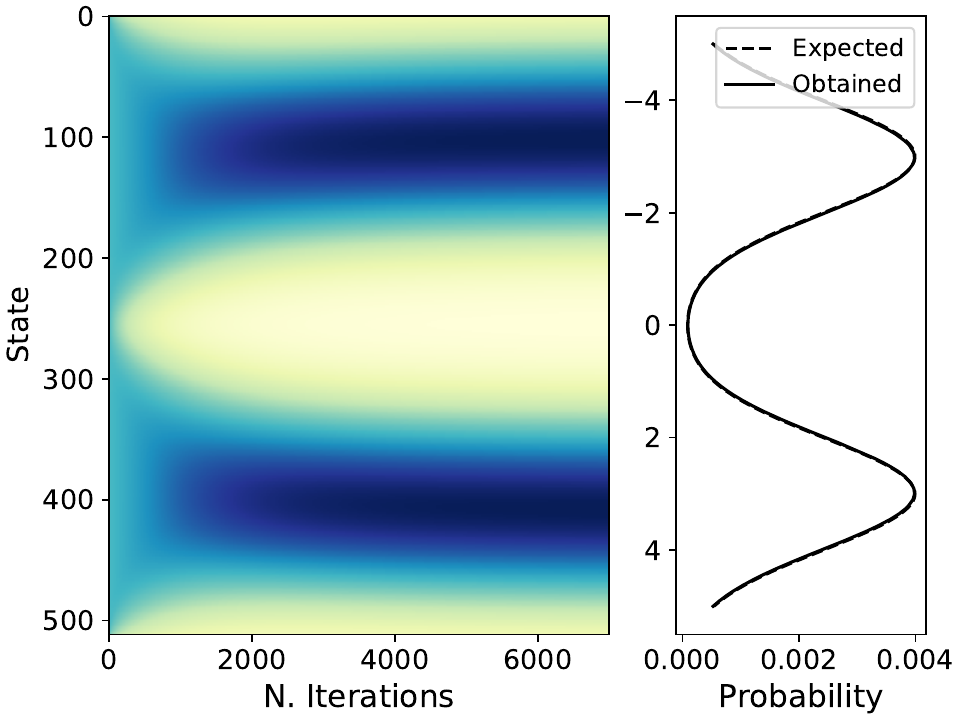}
}
\subfloat[$ |\mathcal{H}_a| = 2^8 $]{
    \label{fig:improve_8}
    \includegraphics[width=0.32\linewidth]{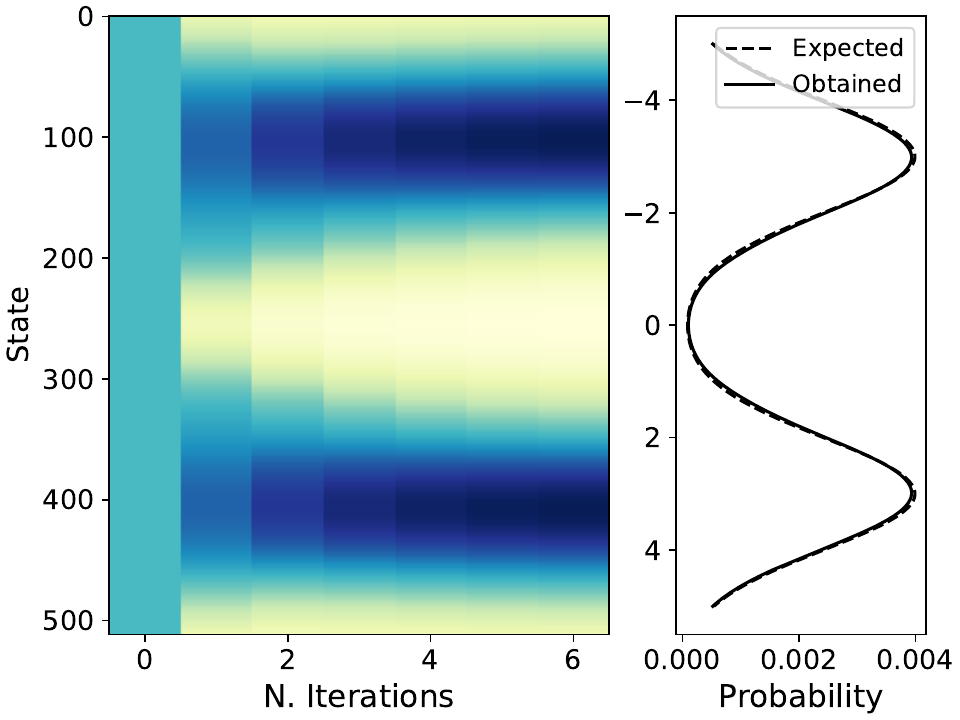}
}
\vspace{0.2cm}
\caption{The figure presents the results of simulations using progressively larger sizes of the register $ \ket{a} $, which enables the evaluation of more candidate moves per iteration when sampling from the mixture of Gaussians $ \mathcal{N}(-3,1) + \mathcal{N}(3,1) $. The subfigures show how increasing the size of $ \ket{a} $ leads to a significant reduction in the number of iterations required to reach convergence. The size of the $ \ket{a} $ register is indicated in the caption of each subfigure (as a power of two, representing the number of qubits).}

\label{fig:improve_figs}
\end{figure*}

Figure~\ref{fig:improve_figs} presents a set of experiments conducted to analyze the effect of register $ \ket{a} $ size on the convergence behavior of the circuit. In all simulations, the target function $ f $ was a mixture of Gaussians of the form $ \mathcal{N}(-3;1) + \mathcal{N}(3;1) $, and the position register $ \ket{x} $ was configured with 9 qubits. In Subfigure~\ref{fig:improve_1}, the baseline configuration is shown, with $ |\mathcal{H}_a| = 2^1 $, where the previously discussed behavior is evident: the number of iterations required for convergence exceeds by far the number of available states in $ \ket{x} $. When the size of $ \ket{a} $ is increased by one qubit (Subfigure \ref{fig:improve_2}), the required number of iterations is reduced by nearly half. This outcome aligns with expectations, as each additional qubit doubles the number of available candidate moves.

Finally, in Subfigure~\ref{fig:improve_8}, where the register $ \ket{a} $ reaches its maximum size for $ |\mathcal{H}_x| = 2^8 $, convergence is achieved with fewer iterations than the number of possible states in $ \ket{x} $. Notably, this reduction in the number of required iterations below the cardinality of the state space, has been consistently observed, independently of the choice of function $ f $ and the size of $ \ket{x} $. This improvement is therefore not limited to the specific case presented in Figure~\ref{fig:improve_figs}.

While the reduction in the number of iterations is significant, it comes at the expense of increased computational resource usage. As discussed in Section~\ref{sec:proposal}, the register $ \ket{a} $ is replaced in each iteration, which means that increasing its size directly multiplies the total number of qubits required by the circuit. This effect is quantified in Equation~\eqref{eq:qubit_increase}, where $ n_a $ denotes the number of qubits in $ \ket{a} $.

At this point, it is natural to question whether the gain in convergence speed justifies the additional qubit cost, especially in the context of current quantum hardware limitations. For instance, in the simulations shown in Subfigures \ref{fig:improve_1} and \ref{fig:improve_8}, and based on Equation~\eqref{eq:qubit_increase}, we find that the first simulation requires 30.027 qubits\footnote{This count includes only the logical qubits forming the registers discussed; ancillary qubits required for circuit implementation are not considered.} to reach convergence after 15.000 iterations. In contrast, using the larger $ \ket{a} $ register configuration allows convergence in just 6 iterations, using only 87 qubits. This represents an effective reduction of over 99.7\% in the total number of qubits required. This behavior has been consistently observed in multiple experiments, with the reduction becoming more pronounced as the size of the register $ \ket{x} $ increases.

\subsection{Comparison with Metropolis-Hastings}
\label{subsec:q_vs_c}

Once the circuit has converged, obtaining a sample is achieved by measuring the state of $ \ket{x} $; register that encodes the target distribution. However, measuring the register destroys the superposition and resets the system, requiring the entire circuit to be rerun to generate each additional sample from the target distribution. This behavior of the circuit is a major disadvantage compared to classical methods.

Let $ T_q $ and $ T_c $ be the number of iterations needed to reach convergence by our quantum circuit and the classical Metropolis-Hastings method, respectively. As mentioned, if the goal is to obtain a total of $ N $ samples from the target distribution, with the quantum method $ N \cdot T_q$ iterations are necessary. In contrast, once converged, the classical methods can generate $ N $ samples in successive iterations, so the sampling time depends on the expression $ T_c + N\cdot T_l $, where $ T_l $ is the number of iterations between samples to reduce the autocorrelation between them and emulate the obtaining of independent samples.

This implies that the quantum circuit would only improve in performance to the classical method when (i) the convergence time of the circuit is $ T_q < T_l $, ensuring a shorter sampling time, and (ii) when the time $ T_q $ is much smaller than the time of convergence of the classical method $ T_c $; that is, $ T_q \lll T_c $; allowing the circuit to get the samples before the convergence of the classical method. In our experiments, we observed that configurations allowing a wider range of movements, such as when  $ |\mathcal{H}_a| = |\mathcal{H}_x|/2 $, lead to a significant reduction in the convergence time $ T_q $, even below the size of the space to be explored. Although these results are promising, they do not constitute a formal proof of convergence or superiority over classical methods.

\begin{figure*}[t]
\centering
\subfloat[Proposed quantum circuit]{
    \label{fig:qmcmc}
    \includegraphics[width=0.49\linewidth]{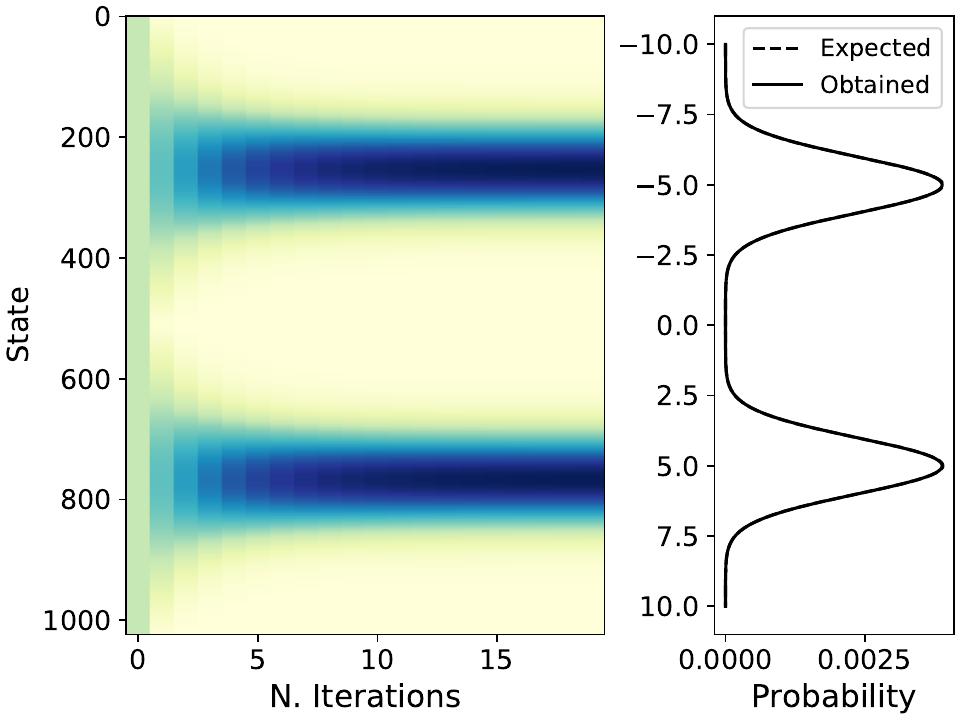}
}
\subfloat[Classical MH method]{
    \label{fig:mh}
    \includegraphics[width=0.49\linewidth]{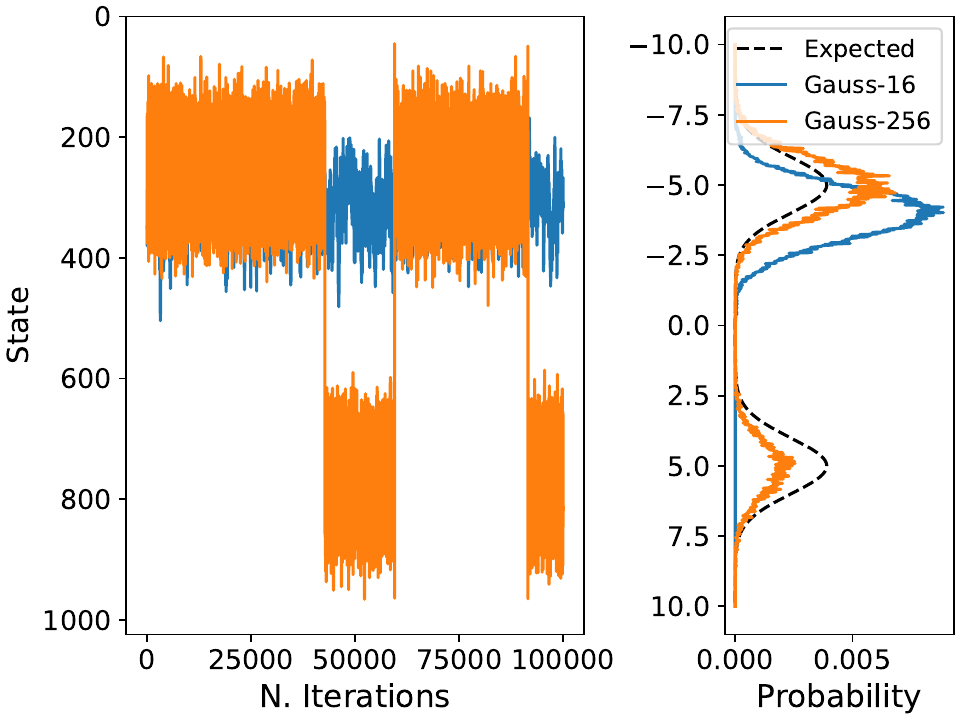}
}
\vspace{0.2cm}
\caption{The figure shows the results obtained from sampling the $ \mathcal{N}(-5,1)+\mathcal{N}(5,1) $ distribution in the interval [-10,10] with a state space size of $ 2^{10} = 1024 $ . The following have been used: (a) the proposed quantum circuit and (b) two Metropolis-Hastings (MH) processes with different generating distributions $ g $. In the classical methods, the generator distributions $ g $ are Gaussian distributions over different amounts of adjacent states; being Gauss-16 (blue line) the chain generated with a Gaussian over the 16 adjacent states as a function $ g $ and Gauss-256 (orange line) the chain generated by a Gaussian over the 256 adjacent states. In the quantum circuit, a size $ \ket{a} = 2^9 $ is used. It can be seen how the circuit converges to the target distribution in as little as 20 iterations; while only the MH process with $ g $ of higher dispersion manages to explore efficiently the space. The MH process with the lowest dispersion in $ g $ gets stuck in one of the modes due to the locality of its movements.}

\label{fig:qmcmc_vs_mh}
\end{figure*}

Figure \ref{fig:qmcmc_vs_mh} shows the results using both our quantum circuit (Subfigure \ref{fig:qmcmc}) and the classical Metropolis-Hastings method with different generator functions $ g $, applied to the same bimodal target. In these experiments, the size of the $ \ket{x} $ register was increased to $ |\mathcal{H}_x| = 2^{10} $ and the search space was [-10,10] in the one dimensional line. The objective function $ f $ to be sampled is the mixture of Gaussians $ \mathcal{N}(-5,1)+\mathcal{N}(5,1) $.

In the classical method, the appropriate choice of the proposal distribution $ g $ is crucial for generating efficient chains. If $ g $ produces trials too close to the current position, the chain may have difficulty exploring the space effectively, increasing the risk of getting caught in one of the modes of the target distribution. Conversely, a distribution $ g $ with greater dispersion allows for proposals that are further away, which favors exploration. However, excessive dispersion can generate trials that are very far away and with a very low probability in $ f $, which increases the probability of rejection and slows down the convergence of the process. A good distribution $ g $ must have the appropriate dispersion to avoid these two effects.

As can be seen in Subfigure \ref{fig:mh}, the chain generated with a $ g $ that only considers the 16 adjacent states (blue lines) cannot efficiently explore the space due to the locality of its movements; getting stuck in one of the modes of the $ f $ distribution. Increasing the dispersion of the distribution to consider all 256 adjacent states (orange lines) allows the generation of trials outside the current mode, allowing the chain to eventually escape the mode. Finding a good $ g $ distribution for each case is not a trivial task and usually resorts to fine-tuning. One of the advantages of our circuit (Subfigure \ref{fig:qmcmc}) is that it does not require fine-tuning to reach convergence. This is because a high-dispersion configuration can always be used and, since trials are accepted and rejected in superposition, the distribution converges in a relatively low number of iterations.

Furthermore, as can be seen in Subfigure \ref{fig:qmcmc}, in the quantum circuit the obtained distribution (solid line) is a very faithful approximation to the function $ f $ (dashed lines), allowing the generation of independent and representative samples of the target distribution. On the other hand, due to the dynamics of the classical method, it focuses only on the high probability regions; therefore, it tends to underestimate or completely eliminate the low probability regions as seen in Subfigure \ref{fig:mh}, specifically at the extremes of space.

In this experiment, the MH method, after 100k iterations, was able to generate a sample of approximately 9000 points (assuming a burn-in of 10\% and a $ T_l $ of 10); while the quantum circuit in that number of iterations can generate up to 5000 totally independent samples. Although MH slightly outperforms our method in this specific low-dimensional example, we expect that in higher-dimensional or more complex distributions the convergence speed of the classical method will be remarkably reduced, increasing the difference between $ T_q $ and $ T_c $. Despite this, one has to take into consideration that: (i) a theoretical study of the circuit properties is needed, especially with regard to the speed of convergence, in order to generalize the result of the experiment to all possible cases, and (ii) the comparison with the Metropolis-Hastings method is very limited, given that there are more classical MCMC methods designed to overcome the problems of the original MH method, improving the sampling capabilities of multimodal and multidimensional distributions. Both, the theoretical study and the comparison with other classical methods are beyond the scope of this paper, although the future work section in the Conclusions section explains in more detail the direction of research on these topics.

\section{Conclusions}
\label{sec:conclusions}

In this paper, we have proposed a modification of the Discrete Quantum Walk (DQW) circuit targeting the Markov Chain Monte Carlo (MCMC) method, encoding the target distribution to be sampled in the register $ \ket{x} $. The results show that the circuit successfully captures the structure of the target distribution $ f $, emulating the behavior of the classical MH algorithm described in Subsection~\ref{subsec:mcmc}. Furthermore, we introduced a modification that enables the evaluation of more trial moves per iteration, thereby enhancing the speed of convergence while reducing the number of qubits required for execution. Finally, we compared our circuit with the classical MH when sampling a multimodal distribution in a one-dimensional space.

However, as pointed out at the end of Subsection \ref{subsec:q_vs_c}, although the results are promising, a theoretical study of the circuit’s properties is required to establish that our method is superior to classical MCMC approaches in all cases. This suggests several lines of research: (i) a formal theoretical analysis of the circuit’s convergence properties, including mixing time, hitting time, and the role of superposition in accelerating convergence compared to classical MCMC, (ii) deriving bounds on the number of iterations required as a function of circuit parameters and the complexity of the target distribution, and (iii) comparison with more advanced classical and quantum MCMC methods, that are better suited for sampling multimodal distributions. 

In relation to the theoretical study, it would be interesting to (iv) determine the optimal register size $ \ket{\text{acc}} $ of our circuit, depending on $ f $, to obtain the most faithful approximation to the target distribution with the smallest possible number of qubits. At last, in all this paper we have only considered a completely symmetrical generator function $ g $ for the trial generation. It is of great relevance to investigate (v) how, following the design of our circuit, to implement any given $ g $ distribution, making the circuit more general and applicable to a wider range of situations.

The operators defined throughout this work remain at a high level of abstraction, describing the circuit in a general manner based on the algorithmic steps of the classical Metropolis–Hastings (MH) method. Therefore, a relevant line of future research (vi) is the decomposition of these operators into their fundamental gates, aiming to reduce the level of abstraction and enable a real implementation. Once achieved, the resulting circuit can be compared with existing quantum methods in terms of circuit depth, number of qubits, and related metrics.

\backmatter

\bmhead{Acknowledgements}
This research is supported by the Basque Government through IKUR-Quantum grant and projects IT1504-22, KK-2024/00030, KK-2024/00068 and KK-2025/00098. It is also supported by the grants PID2023-152390NB-I00 and CNS2023-144315 funded by MICIU/AEI/10.13039/9/501100011033, ``European Union NextGenerationEU/PRTR'' and by ``FEDER funds''. The work at the Center for Theoretical Biological Physics was sponsored by the NSF (Grant PHY-2019745), by the NSF- PHY-2210291, and by the Welch Foundation (Grant C-1792)

\section*{Declarations}

\begin{itemize}
\item Funding\\
The funding is specified in above Acknowledgements section.

\item Conflict of interest\\
The authors did not receive support from any organization that may gain or lose financially through publication of this manuscript.

\item Ethics approval and consent to participate (Not applicable)
\item Consent for publication (Not applicable)
\item Data availability (Not applicable)
\item Materials availability (Not applicable)
\item Code availability\\
The code is available from the corresponding author on reasonable request.

\item Author contribution\\
All authors contributed to the study conception and design.
\end{itemize}


\bibliography{sn-bibliography}

@book{frenkel,
    author = {Frenkel, Daan and Smit, Berend},
    title = {Understanding Molecular Simulation: From Algorithms to Applications},
    year = {1996},
    isbn = {0122673700},
    publisher = {Academic Press, Inc.},
    address = {USA},
    edition = {1st}
}

@article{metropolis,
    author = {Metropolis, Nicholas and Rosenbluth, Arianna W. and et al.},
    title = {Equation of State Calculations by Fast Computing Machines},
    journal = {The Journal of Chemical Physics},
    volume = {21},
    number = {6},
    pages = {1087-1092},
    year = {1953},
    month = {06},
    issn = {0021-9606},
    doi = {10.1063/1.1699114},
    url = {https://doi.org/10.1063/1.1699114}
}

@article{hastings,
    author = {Hastings, W. K.},
    title = {Monte Carlo sampling methods using Markov chains and their applications},
    journal = {Biometrika},
    volume = {57},
    number = {1},
    pages = {97-109},
    year = {1970},
    month = {04},
    issn = {0006-3444},
    doi = {10.1093/biomet/57.1.97},
    url = {https://doi.org/10.1093/biomet/57.1.97}
}

@book{nielsen,
    author = {Nielsen, Michael A. and Chuang, Isaac L.},
    title = {Quantum Computation and Quantum Information},
    publisher = {Cambridge University Press},
    year = {2000},
    isbn = {978-1-107-00217-3},
    address = {UK},
    edition = {10th Anniversary edition}
}

@article{preskill,
    title={Quantum Computing in the {NISQ} era and beyond},
    volume={2},
    ISSN={2521-327X},
    url={http://dx.doi.org/10.22331/q-2018-08-06-79},
    doi={10.22331/q-2018-08-06-79},
    journal={Quantum},
    publisher={Verein zur Forderung des Open Access Publizierens in den Quantenwissenschaften},
    author={Preskill, John},
    year={2018},
    month=aug, pages={79}
}

@article{aharonov,
    title = {Quantum random walks},
    author = {Aharonov, Y. and Davidovich, L. and Zagury, N.},
    journal = {Phys. Rev. A},
    volume = {48},
    issue = {2},
    pages = {1687--1690},
    numpages = {0},
    year = {1993},
    month = {Aug},
    publisher = {American Physical Society},
    doi = {10.1103/PhysRevA.48.1687},
    url = {https://link.aps.org/doi/10.1103/PhysRevA.48.1687}
}

@article{ambainis_qw1d,
    author = {Ambainis, Andris},
    title = {Quantum Walk Algorithm for Element Distinctness},
    year = {2007},
    issue_date = {April 2007},
    publisher = {Society for Industrial and Applied Mathematics},
    address = {USA},
    volume = {37},
    number = {1},
    issn = {0097-5397},
    url = {https://doi.org/10.1137/S0097539705447311},
    doi = {10.1137/S0097539705447311},
    journal = {SIAM J. Comput.},
    month = apr,
    pages = {210–239},
    numpages = {30}
}

@inproceedings{szegedy,
    author={Szegedy, M.},
    booktitle={45th Annual IEEE Symposium on Foundations of Computer Science}, 
    title={Quantum speed-up of Markov chain based algorithms}, 
    year={2004},
    volume={},
    number={},
    pages={32-41},
    doi={10.1109/FOCS.2004.53}
}

@article{child_universal,
    title = {Universal Computation by Quantum Walk},
    author = {Childs, Andrew M.},
    journal = {Phys. Rev. Lett.},
    volume = {102},
    issue = {18},
    pages = {180501},
    numpages = {4},
    year = {2009},
    month = {May},
    publisher = {American Physical Society},
    doi = {10.1103/PhysRevLett.102.180501},
    url = {https://link.aps.org/doi/10.1103/PhysRevLett.102.180501}
}

@article{zahringer,
    title = {Realization of a Quantum Walk with One and Two Trapped Ions},
    author = {Z\"ahringer, F. and Kirchmair, G. and et al.},
    journal = {Phys. Rev. Lett.},
    volume = {104},
    issue = {10},
    pages = {100503},
    numpages = {4},
    year = {2010},
    month = {Mar},
    publisher = {American Physical Society},
    doi = {10.1103/PhysRevLett.104.100503},
    url = {https://link.aps.org/doi/10.1103/PhysRevLett.104.100503}
}

@article{schreiber,
    author = {Andreas Schreiber and Aurél Gábris and et al},
    title = {A 2D Quantum Walk Simulation of Two-Particle Dynamics},
    journal = {Science},
    volume = {336},
    number = {6077},
    pages = {55-58},
    year = {2012},
    doi = {10.1126/science.1218448},
    url = {https://www.science.org/doi/abs/10.1126/science.1218448}
}

@article{peruzzo,
    author = {Alberto Peruzzo and Mirko Lobino and et al.},
    title = {Quantum Walks of Correlated Photons},
    journal = {Science},
    volume = {329},
    number = {5998},
    pages = {1500-1503},
    year = {2010},
    doi = {10.1126/science.1193515},
    url = {https://www.science.org/doi/abs/10.1126/science.1193515}
}

@article{gibbs,
    author={Geman, Stuart and Geman, Donald},
    journal={IEEE Transactions on Pattern Analysis and Machine Intelligence}, 
    title={Stochastic Relaxation, Gibbs Distributions, and the Bayesian Restoration of Images}, 
    year={1984},
    volume={PAMI-6},
    number={6},
    pages={721-741},
    doi={10.1109/TPAMI.1984.4767596}
}

@article{pt_theory,
    author ="Earl, David J. and Deem, Michael W.",
    title  ="Parallel tempering: Theory{,} applications{,} and new perspectives",
    journal  ="Phys. Chem. Chem. Phys.",
    year  ="2005",
    volume  ="7",
    issue  ="23",
    pages  ="3910-3916",
    publisher  ="The Royal Society of Chemistry",
    doi  ="10.1039/B509983H",
    url  ="http://dx.doi.org/10.1039/B509983H"
}

@article{pt_deri_adaptativept,
    author = {Błażej Miasojedow and Eric Moulines and Matti Vihola},
    title = {An Adaptive Parallel Tempering Algorithm},
    journal = {Journal of Computational and Graphical Statistics},
    volume = {22},
    number = {3},
    pages = {649--664},
    year = {2013},
    publisher = {ASA Website},
    doi = {10.1080/10618600.2013.778779},
    url = {https://doi.org/10.1080/10618600.2013.778779}
}

@misc{tucci,
      author={Robert R. Tucci},
      title={Quantum Gibbs Sampling Using Szegedy Operators}, 
      year={2010},
      archivePrefix={arXiv},
      primaryClass={quant-ph},
      url={https://arxiv.org/abs/0910.1647}, 
}

@article{qaoa_betterconvergence,
  title={Warm-started QAOA with custom mixers provably converges and computationally beats goemans-williamson's max-cut at low circuit depths},
  author={Tate, Reuben and Moondra, Jai and Gard, Bryan and Mohler, Greg and Gupta, Swati},
  journal={Quantum},
  volume={7},
  pages={1121},
  year={2023},
  publisher={Verein zur F{\"o}rderung des Open Access Publizierens in den Quantenwissenschaften}
}

@article{qaoa_customoperator,
  title={From the quantum approximate optimization algorithm to a quantum alternating operator ansatz},
  author={Hadfield, Stuart and Wang, Zhihui and O’gorman, Bryan and Rieffel, Eleanor G and Venturelli, Davide and Biswas, Rupak},
  journal={Algorithms},
  volume={12},
  number={2},
  pages={34},
  year={2019},
  publisher={MDPI}
}

@article{qaoa_thermaldistro,
  title={Connection between single-layer quantum approximate optimization algorithm interferometry and thermal distribution sampling},
  author={D{\'\i}ez-Valle, Pablo and Porras, Diego and Garc{\'\i}a-Ripoll, Juan Jos{\'e}},
  journal={Frontiers in Quantum Science and Technology},
  volume={3},
  pages={1321264},
  year={2024},
  publisher={Frontiers Media SA}
}

@article{qa_dwave,
  title={Quantum annealing with Markov chain Monte Carlo simulations and D-wave quantum computers},
  author={Wang, Yazhen and Wu, Shang and Zou, Jian},
  journal={Statistical Science},
  pages={362--398},
  year={2016},
  publisher={JSTOR}
}

@article{qa_mcmc,
  title={Quantum annealing enhanced Markov-Chain Monte Carlo},
  author={Arai, Shunta and Kadowaki, Tadashi},
  journal={Scientific Reports},
  volume={15},
  number={1},
  pages={21427},
  year={2025},
  publisher={Nature Publishing Group UK London}
}

@article{hybrid_mcmc,
  title = {Quantum-enhanced Markov chain Monte Carlo},
  author = {Layden, D. and Mazzola, G. and Mishmash, R.V.},
  journal = {Nature},
  volume = {619},
  issue = {4},
  pages = {282-287},
  numpages = {8},
  year = {2023},
  publisher = {American Physical Society},
  doi = {https://doi.org/10.1038/s41586-023-06095-4},
}

@article{hybrid_moremcmc,
  title={Quantum-enhanced Markov chain Monte Carlo for systems larger than a quantum computer},
  author={Ferguson, Stuart and Wallden, Petros},
  journal={Physical Review Research},
  volume={7},
  number={1},
  pages={013231},
  year={2025},
  publisher={APS}
}

\end{document}